\documentclass[10pt, final, journal, letterpaper, twocolumn]{IEEEtran}
\usepackage{amsmath}
\usepackage{amssymb}
\usepackage{amsfonts}
\usepackage{graphicx}
\usepackage{epsfig}
\usepackage{subfigure}
\usepackage{psfrag}
\usepackage{cite}
\usepackage{latexsym}
\usepackage{url}
\usepackage{color}
\usepackage{algorithm}
\usepackage{algorithmic}
\usepackage{bbm}
\usepackage{multirow}
\usepackage{array}

\PassOptionsToPackage{bookmarks={false}}{hyperref}

\newtheorem{remark}{\underline{Remark}}[section]
\newcommand{\mv}[1]{\mbox{\boldmath{$ #1 $}}}

\begin{document}
	
	\title{UAV-Enabled Data Collection for Wireless Sensor Networks with Distributed Beamforming}	
\author{Tianxin Feng, Lifeng Xie, Jianping Yao, and Jie Xu
\thanks{This paper will be present in part at the IEEE International Conference on Communications Workshops (ICC Workshops) on ``Integrating UAVs into 5G and Beyond", Virtual Conference, June 7-11, 2020 \cite{FengOutage2020}.}
\thanks{T. Feng, L. Xie, and J. Yao are with the School of Information Engineering, Guangdong University of Technology, Guangzhou 510006, China (e-mail:~ftx.gdut@gmail.com,~lifengxie@mail2.gdut.edu.cn,~and yaojp@gdut.edu.cn). J. Yao is the corresponding author.}
\thanks{J. Xu is with the Future Network of Intelligence Institute (FNii) and the School of Science and Engineering, The Chinese University of Hong Kong, Shenzhen, Shenzhen 518172, China (e-mail: xujie@cuhk.edu.cn).}
}
\maketitle
	\begin{abstract}		
	This paper studies an unmanned aerial vehicle (UAV)-enabled wireless sensor network, in which one UAV flies in the sky to collect the data transmitted from a set of ground nodes (GNs) via distributed beamforming. We consider two scenarios with delay-tolerant and delay-sensitive applications, in which the GNs send the common/shared messages to the UAV via adaptive- and fixed-rate transmissions, respectively. For the two scenarios, we aim to maximize the average data-rate throughput and minimize the transmission outage probability, respectively, by jointly optimizing the UAV's trajectory design and the GNs' transmit power allocation over time, subject to the UAV's flight speed constraints and the GNs' individual average power constraints. However, the two formulated problems are both non-convex and thus generally difficult to be optimally solved. To tackle this issue, we first consider the relaxed problems in the ideal case with the UAV's flight speed constraints ignored, for which the well-structured optimal solutions are obtained to reveal the fundamental performance upper bounds. It is shown that for the two approximate problems, the optimal trajectory solutions have the same multi-location-hovering structure, but with different optimal power allocation strategies. Next, for the general problems with the UAV's flight speed constraints considered, we propose efficient algorithms to obtain high-quality solutions by using the techniques from convex optimization and approximation. Finally, numerical results show that our proposed designs significantly outperform other benchmark schemes, in terms of the achieved data-rate throughput and outage probability under the two scenarios. It is also observed that when the mission period becomes sufficiently long, our proposed designs approach the performance upper bounds when the UAV's flight speed constraints are ignored.
	\end{abstract}		
	\section{Introduction}\label{sec:int}
	
	\IEEEPARstart{U}nmanned aerial vehicles (UAVs) or drones are expected to have a lot of applications in beyond-fifth-generation (B5G) and sixth-generation (6G) wireless networks as dedicatedly deployed aerial wireless platforms (such as aerial base stations (BSs) \cite{MozaffariBeyond2019,LiPlacement2018,Alzenad3-D2017,ZhongSecure2019,YaoSecrecy2019,BanagarPerformance2020,SharmaRandom2019}, cellular-connected users \cite{AmerMobility2020,ZhangTrajectory2019}, energy transmitters (ETs) \cite{XieThroughput2019,XuUAV2018,XieCommon2020}, relays \cite{ChenEfficient2020,ChenMultiple2018}, and mobile edge computing (MEC) servers \cite{ZhouMobile2020,HuJoint2019}).
Among others, there has been an upsurge of interest in using UAVs as aerial data collectors (or fusion centers) to collect data from large-scale wireless sensor networks (WSNs).
In the upcoming Internet of Things (IoT) era, WSNs have been widely deployed for applications such as surveillance and environmental, agricultural, and traffic monitoring \cite{LiEnergy2019,lei,Jawad,Nellore}, by collecting, e.g., geographical and environmental information, as well as images and videos.
In practice, these applications might be implemented in the remote areas (e.g., rural macro) or in emergency situations (e.g., after an earthquake or in the sea), such that employing ground BSs for data collections may become difficult if not infeasible \cite{AlmasoudEnergy2018}.
How to collect the data to the BSs in a fast and reliable manner is one of the key challenges faced in the design of WSNs.
Different from the conventional designs using on-ground fusion centers for data collection, the UAVs in the sky can exploit the fully-controllable mobility in the three-dimensional (3D) space to fly close to the IoT devices for collecting data more efficiently, and then convey the data to the target BSs. UAVs can also leverage the strong line-of-sight (LoS) ground-to-air (G2A) channels for increasing the communication quality.
	
	In the literature, there have been a handful of prior works studying the UAV-enabled data collection \cite{GongFlight2018,LiJoint2019,WangEnergy2019,ZhanEnergy2018,ZhangMulti2020,You3D2019,Li2020}, in which the UAV's trajectory is designed for enhancing the system performance.
	For example, the authors in \cite{GongFlight2018} and \cite{LiJoint2019} jointly designed the UAV's flight trajectory and wireless resource allocation/scheduling to minimize the mission completion time, in the scenarios when the sensors are deployed in the one-dimensional (1D) and two-dimensional (2D) spaces, respectively. The authors in \cite{WangEnergy2019} and \cite{ZhanEnergy2018} optimized the UAV's trajectory and the sensors' transmission/wakeup scheduling, in order to maximize the energy efficiency of the WSNs while ensuring the collected data amounts from sensors.
	The authors in \cite{ZhangMulti2020} jointly designed the sensors' transmission scheduling, power allocations, and UAV's trajectory to maximize the minimum data collection rate from the ground sensors to a multi-antenna UAV.
	Furthermore, \cite{You3D2019} exploited the UAV's 3D trajectory optimization for maximizing the minimum average rate for data collection, by considering angle-dependent Rician fading channels. In addition, \cite{Li2020} characterized the fundamental rate limits of UAV-enabled multiple access channels (MAC) for data collection in a simplified scenario with linearly deployed sensors on the ground.
	In these prior works, the authors considered the adaptive-rate transmission at the sensors, such that the sensors on the ground can adaptively adjust their transmission rate based on the wireless channel fluctuations due to the mobility of the UAVs.
	Furthermore, these prior works assumed that the on-ground devices (or sensors) send independent messages to the UAV under different multiple access techniques.

Nowadays with technical advancements, the sensors are becoming more powerful with more advanced signal processing capabilities to support new applications, such as distributed beamforming.
In contrast to the communicating independently, distributed beamforming has been recognized as a promising technique to enhance the data rate and energy efficiency in WSNs (see, e.g., \cite{1,Z,Haro} and the references therein), in which a large number of sensors are enabled to coordinate in transmitting common or shared messages to a fusion center (the UAV of our interest).
	By properly controlling the phases, the signals transmitted from different sensors can be coherently combined at the fusion center, thus increasing the communication range and enhancing the energy efficiency via exploiting the distributed beamforming gain \cite{1}.
For example, the authors in \cite{Z} considered a wireless powered communication networks system, in which the sensors first harvest energy from dedicated ETs and then transmit information to a fixed access point (AP), to enhance the transmission performance via designing the distributed beamforming.
	The authors in \cite{Haro} designed the distributed beamforming in order to maximize the network
	lifetime under the requirement of a pre-specified quality of service.
	In these prior works, the authors assumed that the fusion centers are fixed on the ground. By contrast, under the mobile fusion center deployed at a UAV of our interest, how to jointly design the UAV's trajectory and the wireless resource allocation for improving the data collection performance is a new problem that has not been investigated yet.
	
Motivated by this, this paper focuses on a new UAV-enabled data collection system with distributed beamforming, in which the UAV collects data from multiple single-antenna ground nodes (GNs) via the distributed beamforming. The GNs have powerful computing and communication capabilities and can be the headers of the sensors or nodes with sensing capability themselves which are deployed in the same swarm and interconnected with device-to-device (D2D) communications. In each swarm, due to the random errors such as quantization noise and wireless channel distortion, the sensing data synchronization scheme is adopted, so that the transmitted messages by all the GNs are identical. For example, all GNs can broadcast the collected sensing data to each other until the data is synchronized. Notice that when the UAV flies over the swarm of GNs, the UAV will broadcast the reference signals periodically so that the GNs perform carrier synchronization and channel estimation \cite{Barriac,m2}.

Different from the existing works focusing on the adaptive-rate transmissions at the ground devices, we consider two scenarios with the adaptive-rate and fixed-rate transmissions. These two scenarios may correspond to the delay-tolerant applications (e.g., for delay-insensitive measurement information delivery) and the delay-sensitive applications (e.g., for real-time video delivery), respectively. For the two scenarios, our objectives are to maximize the average data-rate throughput and minimize the transmission outage probability, respectively, by jointly optimizing the UAV's trajectory design and the GNs' transmit power allocation over time, subject to the UAV's flight speed constraints and the GNs' individual average power constraints.
However, due to the infinite number of optimization variables for the GNs' power allocation and UAV's trajectory over continuous time, how to jointly optimize them is a difficult problem.

%{\small
%		\begin{table*}[!t]
%			\centering
%				\caption{List of Notations}\label{notations}
%				\begin{tabular}{lll}
%					\hline
%					$K$: Number of GNs&$\mathcal{K}$: Set of GNs&$T$: Mission duration\\
%					
%					$\mathcal{T}$: Mission period&$\boldsymbol{s}_k$: Horizontal location of GN $k$&$N/\tilde{N}$: Number of time slots for $(\mathtt{P1})/(\mathtt{P2})$\\
%					
%					$\mathcal{N}/\tilde{\mathcal{N}}$: Set of time slots for $(\mathtt{P1})/(\mathtt{P2})$&$\boldsymbol{q}$: UAV's location&$\boldsymbol{q}_I/\boldsymbol{q}_F$: UAV's initial/final location\\
%					
%					$V_{\rm max}$: UAV's maximum flight speed&$\mathcal{Z}$: UAV's desirable flight region&$d_{k}$: Distance between UAV and GN $k$\\
%					
%					$\beta_0$: Channel power gain at reference $d_0$&$H$: UAV's fixed altitude&$h_k$: Channel coefficient from GN $k$ to UAV\\
%					
%					$\alpha$: Path loss exponent&$P_k$: GN $k$'s transmit power&$\psi_{k}$: Channel phase shift from GN $k$ to UAV\\
%					
%					$\upsilon$: AWGN&$\sigma^2$: AWGN power&$\delta/\tilde{\delta}$: Discretized duration for $(\mathtt{P1})/(\mathtt{P2})$\\
%					
%					$\varphi_{k}$: GN $k$'s signal phase&$r$: Data-rate throughput&$P^{\rm ave}_k$: GN $k$'s maximum average power\\
%					
%					$\gamma_{\rm min}$: SNR threshold&$O$: Outage probability&${\boldsymbol{w}}$: Distributed beamforming vector\\
%					
%					$\pi$: SNR order permutation over $\mathcal {\tilde{N}}$&${\boldsymbol{h}}$: Combined channel vector&$\mathcal{N}'$: Subset of $\mathcal {\tilde{N}}$ with highest SNR values\\			
%					\hline
%			\end{tabular}
%		\end{table*}
%	}

		\begin{table}[!t]
			\centering
				\caption{List of Notations}\label{notations}
				\begin{tabular}{ll}
					\hline
$K$                                    &Number of GNs\\
$\mathcal{K}$                          &Set of GNs\\
$T$                                    &Mission duration\\
$\mathcal{T}$                          &Mission period\\
$\boldsymbol{s}_k$                     &Horizontal location of GN $k$\\
$N/\tilde{N}$                          &Number of time slots for $(\mathtt{P1})/(\mathtt{P2})$\\
$\mathcal{N}/\tilde{\mathcal{N}}$      &Set of time slots for $(\mathtt{P1})/(\mathtt{P2})$\\
$\boldsymbol{q}$                       &UAV's horizontal location\\
$\boldsymbol{q}_I/\boldsymbol{q}_F$    &UAV's horizontal initial/final location\\
$V_{\rm max}$                          &UAV's maximum flight speed\\
$\mathcal{Z}$                          &UAV's desirable flight region\\
$d_{k}$                                &Distance between UAV and GN $k$\\
$\beta_0$                              &Channel power gain at reference $d_0$\\
$H$                                    &UAV's fixed altitude\\
$h_k$                                  &Channel coefficient from GN $k$ to UAV\\
$\alpha$                               &Path loss exponent\\
$P_k$                                  &GN $k$'s transmit power\\
$\psi_{k}$                             &Channel phase shift from GN $k$ to UAV\\
$\upsilon$                             &AWGN\\
$\sigma^2$                             &AWGN power\\
$\delta/\tilde{\delta}$                &Discretized duration for $(\mathtt{P1})/(\mathtt{P2})$\\
$\varphi_{k}$                          &GN $k$'s signal phase\\
$r$                                    &Data-rate throughput\\
$P^{\rm ave}_k$                        &GN $k$'s maximum average power\\
$\gamma_{\rm min}$                     &SNR threshold\\
$O$                                    &Outage probability\\
${\boldsymbol{w}}$                     &Distributed beamforming vector\\
$\pi$                                  &SNR order permutation over $\mathcal {\tilde{N}}$\\
${\boldsymbol{h}}$                     &Combined channel vector\\
$\mathcal{N}'$                         &Subset of $\mathcal {\tilde{N}}$ with highest SNR values\\			
					\hline
			\end{tabular}
		\end{table}
	
	To deal with this issue, we first consider the relaxed problems in the ideal case without considering the UAV's flight speed constraints, for which the well-structured optimal solutions are obtained via the Lagrange duality method to reveal the fundamental performance upper bounds.
It is observed that for the two scenarios, the optimal trajectory solutions follow the same multi-location-hovering structure, but the optimal power allocation solutions are distinct.
In particular, in the first scenario for rate maximization, the GNs transmit their messages based on the water-filling-like power allocation over time. However, in the second scenario for outage probability minimization, the GNs adopt an on-off power allocation over time, where the GNs may remain silent in the outage status when the wireless channels become bad, such that the transmit power can be reserved for non-outage transmission at other time instants.
	
Next, motivated by the obtained optimal trajectories for the above special problems, we propose	efficient approaches to obtain high-quality solutions to the general problems with the UAV's flight speed constraints considered, by using techniques from convex optimization and	approximation. In the proposed approaches, we solve a series of approximated convex optimization problems to update the UAV's flight trajectories and the GNs' power allocations towards efficient solutions.

Finally, we provide numerical simulations to validate the effectiveness of our proposed schemes.
It is shown that our proposed designs significantly outperform the benchmark schemes in terms of the achieved data-rate throughput and outage probability under the two scenarios.
It is also shown that when the communication duration becomes sufficiently long, the proposed designs approach the performance upper bounds achieved when the UAV's flight speed constraints are ignored.
	
	The remainder of this paper is organized as follows. Section \ref{sec:system} introduces the system model of our considered UAV-enabled data collection system with distributed beamforming. Section \ref{s4} solves the average data-rate throughput maximization problem in the delay-tolerant application scenario. Section \ref{s5} solves the outage probability minimization problem in the delay-sensitive application scenario. Section \ref{s7} presents numerical results. Finally, Section \ref{s11} concludes this paper.

\section{System Model and Problem Formulation}\label{sec:system}
	We consider a UAV-enabled data collection system, in which one single-antenna UAV acts as an aerial mobile data collector to periodically collect data from a set of $\mathcal K\triangleq \{1,\ldots,K\}$ single-antenna GNs. We assume that all the GNs collaborate as a swarm to transmit common/shared messages towards the UAV with distributed beamforming employed. It is assumed that each GN $k\in\mathcal K$ is deployed at a fixed location $(x_k,y_k,0)$ on the ground in the 3D Cartesian coordinate system. For notational convenience, let $\boldsymbol{s}_k=(x_k,y_k)$ denote the horizontal location of GN $k\in\mathcal K$, which is assumed to be known by the UAV {\it a-priori} to facilitate the trajectory design.\footnote{Since the GNs are active, the UAV can estimate the GNs' locations according to the signal angle-of-arrival (AOA) or the received signal strength (RSS). Furthermore, due to the high altitude of the UAV, a UAV can locate the GNs by using its optical camera and synthetic aperture radar with intelligent image, video processing, and pattern recognition \cite{ShangUnmanned2019}.} For ease of reading, we summarize the main notations in this paper in Table \ref{notations}.
			
	We focus on one particular mission period of the UAV with finite duration $T$ in second (s), denoted by $\mathcal{T}\triangleq [0,T]$. The UAV is assumed to fly at a fixed altitude $H$, with the time-varying horizontal location $\boldsymbol{q}(t)=(x(t),y(t))$ for any time instant $t\in\mathcal{T}$.
	Suppose that $\boldsymbol{q}_{\rm I}=(x_I,y_I)$ and $\boldsymbol{q}_{\rm F}=(x_F,y_F)$ denote the UAV's horizontal initial and final locations, respectively.
 Let $V_{\rm max}$ denote the UAV's maximum flight speed. Thus, we
	have
	\begin{align}
		\dot{x}^2(t)+\dot{y}^2(t)\leq V^2_{\rm max}, \forall t\in\mathcal{T},\label{speed}\\
		\boldsymbol{q}(0)=\boldsymbol{q}_{\rm I}, \boldsymbol{q}(T)=\boldsymbol{q}_{\rm F},\label{location}
	\end{align}
	where $\dot{x}(t)$ and $\dot{y}(t)$ denote the first-derivatives of $x(t)$ and $y(t)$ with respect to $t$, respectively.
	We denote the region $\mathcal{Z}\triangleq[\underline{x},\bar{x}]\times[\underline{y},\bar{y}]$ as the UAV's desirable flight region in the horizontal plane,
	where $\underline{x}=\min(\{x_{k},k\in \mathcal{K}\}\cup \{x_I,x_F\})$, $\bar{x}=\max(\{x_{k},k\in \mathcal{K}\}\cup \{x_I,x_F\})$, $\underline{y}=\min(\{y_{k},k\in \mathcal{K}\}\cup \{y_I,y_F\})$, and $\bar{y}=\max(\{y_{k},k\in \mathcal{K}\}\cup \{y_I,y_F\})$.
	We also assume that the UAV's mission duration $T$ satisfies $T\geq \|\boldsymbol{q}_{\rm F}-\boldsymbol{q}_{\rm I}\|/V_{\rm max}$, in order for the trajectory from the initial to final locations to be feasible.
	Accordingly, the distance between the UAV and GN $k\in\mathcal K$ at any time instant $t\in\mathcal T$ is given by
	\begin{align}
		d_{k}(\boldsymbol{q}(t))=\sqrt{\|\boldsymbol{q}(t)-\boldsymbol{s}_k\|^2+H^2}.\nonumber
	\end{align}
According to the practically measured results in \cite{LinThe2018} and the channel models introduced in Third Generation Partnership Project (3GPP) TR 36.777 \cite{3GPP}, we assume that the UAV's altitude $H$ is sufficiently high so that the G2A channels from the GNs to the UAV are LoS dominated with the path loss exponent $\alpha\in[2,4]$. Therefore, we consider a channel model with LoS path loss together with random phases, which may come from devices' circuits, environmental factors, and UAV's fluctuation and jittering as introduced in \cite{DabiriAnalytical2020,BanagarImpact}. Consequently, the channel coefficient between the UAV and GN $k\in\mathcal K$ at any time instant $t\in\mathcal T$ is given by
	\begin{align}
		{h}_{k}(\boldsymbol{q}(t))=\sqrt{\beta_0d_{k}^{-\alpha}(\boldsymbol{q}(t))}e^{j\psi_{k}(t)},\nonumber
	\end{align}
	where $\beta_0$ denotes the channel power gain at the reference distance of $d_0=1$ m, $\psi_{k}(t)$ denotes the channel phase shift at any time instant $t\in\mathcal T$ \cite{1}.\footnote{At the beginning of each time slot, the GNs can overhear the transmitted signals (particularly pilots) from the UAV, perform channel estimation, quantize the estimated channel phase shift online with the traditional channel estimation methods \cite{ZengOptimized2015}. Note that no channel estimation at the UAV and no feedback operation at the GNs are required.}
	
	In particular, we consider that all the GNs collaborate as a swarm to transmit a common message $s$, which is a circularly symmetric complex Gaussian (CSCG) random variable with zero mean and unit variance (i.e., $s\sim\mathcal{CN}(0,1)$).\footnote{For the common message obtaining, the ground IoT devices (e.g., monitoring sensors) first generate the messages (e.g., physical phenomenon) and send them to the GN nearby. Then, the GN summarizes and processes the received messages as the common messages. After that, the common messages are shared to other GNs by D2D communications. As the GNs are generally located in a very short distance, the energy consumed for data sharing is negligible as compared to the energy consumed for transmission towards the UAV.}
	At any time instant $t\in\mathcal T$, the transmit signal of GN $k\in\mathcal K$ is $\sqrt{P_{k}(t)}e^{j\varphi_{k}(t)}s$, where $P_k(t)\ge 0$ and $\varphi_{k}(t)\in[-\pi,\pi]$ denote GN $k$'s transmit power and signal phase, respectively. Suppose that each GN $k\in\mathcal K$ is subject to a maximum average power budget $P^{\rm ave}_k$. Therefore, the average transmit power constraint for each GN $k$ is given by
	\begin{align}
		\frac{1}{T}\int_{0}^T P_k(t){\rm d}t\leq P^{\rm ave}_k,\forall k\in \mathcal{K}.\label{10202130}
	\end{align}
	Then, the received signal at the UAV at any time instant $t\in \mathcal{T}$ is given by
	\begin{align}
		y(t)=\sum_{k=1}^K\sqrt{P_{k}(t)\beta_0d_{k}^{-\alpha}(\boldsymbol{q}(t))}e^{j(\varphi_{k}(t)+\psi_{k}(t))}s+\upsilon.\nonumber
	\end{align}
	Here, $\upsilon$ denotes the additive white Gaussian noise (AWGN) at the UAV's information receiver, which is a CSCG random variable with zero mean and variance $\sigma^2$ (i.e., $\upsilon\sim\mathcal{CN}(0,\sigma^2)$). Since the GNs can estimate the channel phase shift online, we design the signal phase as $\varphi_{k}(t)=-\psi_{k}(t)$ online to achieve the maximum received signal power at the UAV. After obtaining the locations of the GNs by the methods introduced before, we can solve our proposed problems to design the GNs' transmit powers and the UAV's trajectory offline, thus computing the signal-to-noise ratio (SNR). As a result, the received SNR by the UAV at any time instant $t\in \mathcal{T}$ is given by
	\begin{align}
			{\tt SNR}(\boldsymbol{q}(t),\{P_k(t)\})=\frac{\left(\sum_{k=1}^K\sqrt{P_{k}(t)\beta_0d_{k}^{-\alpha}(\boldsymbol{q}(t))}\right)^2}{\sigma^2}.\label{06051502}
	\end{align}
Consequently, the data-rate throughput from the $K$ GNs to the UAV in bits/second/Hertz (bps/Hz) at time instant $t\in \mathcal{T}$ is given by
	\begin{align}
		r(\boldsymbol{q}(t),\{P_k(t)\})=\log_2\left(1+{{\tt SNR}\left(\boldsymbol{q}(t),\{P_k(t)\}\right)}\right).\label{12251047}
	\end{align}
	
	In the following, we will formulate the optimization problems for rate maximization in the delay-tolerant application scenario and outage probability minimization in the delay-sensitive application scenario, respectively.
	
	\subsection{Rate Maximization in Delay-Tolerant Application Scenario}
	In the delay-tolerant application scenario, we assume that the GNs can adaptively adjust the communication rate based on channel variations due to the time-varying locations of the UAV. In this case, the average or ergodic data-rate throughput is used as the performance metric. According to (\ref{12251047}), the average data-rate throughput from $K$ GNs to the UAV over the whole duration $T$ in bps/Hz is given by
	\begin{align}
		R(\{\boldsymbol{q}(t),P_k(t)\})=\frac{1}{T}\int_{0}^T\log_2\left(1+{{\tt SNR}\left(\boldsymbol{q}(t),\{P_k(t)\}\right)}\right){\rm d}t.\nonumber
	\end{align}
	
	Our objective is to maximize the average data-rate throughput $R(\{\boldsymbol{q}(t),P_k(t)\})$, by jointly optimizing the UAV's trajectory $\{\boldsymbol{q}(t)\}$ and GNs' power allocation $\{P_k(t)\}$ over time, subject to the UAV's flight speed constraints in (\ref{speed}), the UAV's initial and final locations constraints in (\ref{location}), and the GNs' average transmit power constraints in (\ref{10202130}). Consequently, the average data-rate throughput maximization problem is formulated as
	\begin{align}
		(\mathtt{P1}):~\max\limits_{\{\boldsymbol{q}(t)\},\{P_k(t)\geq0\}}~& R(\{\boldsymbol{q}(t),P_k(t)\}) \nonumber\\
		\mathrm{s.t.}~&(\ref{speed}),~(\ref{location}),~\text{and}~(\ref{10202130}).\nonumber
	\end{align}
	It is worth noting that the objective function of problem $(\mathtt{P1})$ is non-concave, due to the complicated data-rate throughput expression with respect to coupled variables $\boldsymbol{q}(t)$'s and $P_k(t)$'s. Moreover, problem $(\mathtt{P1})$ contains an infinite number of optimization variables over continuous time. As a result, problem $(\mathtt{P1})$ is difficult to be solved optimally. We will deal with this issue in Section \ref{s4}.
	
	\subsection{Outage Probability Minimization in Delay-Sensitive Application Scenario}
	In the delay-sensitive application scenario, we assume that the GNs use a fixed transmission rate for delivering the delay-sensitive information.
	In order for the UAV to successfully decode the message (with fixed rate) at any given time instant, the received SNR must be no smaller than a certain threshold $\gamma_{\rm min}$. In this case, the transmission outage occurs if the received SNR at the UAV falls below $\gamma_{\rm min}$.
	Therefore, we use the following indicator function to indicate the transmission outage at any time instant $t\in \mathcal{T}$.
	\begin{align}
		\mathbbm{1}({\tt SNR}(\boldsymbol{q}(t),\{P_k(t)\}))=\left\{
		\begin{array}{ll}
			1,& {\tt SNR}(\boldsymbol{q}(t),\{P_k(t)\})<\gamma_{\rm min},\\
			0,& {\tt SNR}(\boldsymbol{q}(t),\{P_k(t)\})\geq\gamma_{\rm min}.
		\end{array}
		\right.\nonumber
	\end{align}
	Accordingly, we define the outage probability as the probability that the transmission is in outage over the whole duration $T$, which is expressed as
	\begin{align}
		O(\{\boldsymbol{q}(t),P_k(t)\})=\frac{1}{T}\int_{0}^T\mathbbm{1}({\tt SNR}(\boldsymbol{q}(t),\{P_k(t)\})){\rm d}t.\nonumber
	\end{align}
	
	Our objective is to minimize the outage probability $O(\{\boldsymbol{q}(t),P_k(t)\})$, by jointly optimizing the UAV's trajectory $\{\boldsymbol{q}(t)\}$ and GNs' power allocation $\{P_k(t)\}$ over time, subject to the UAV's flight speed constraints in (\ref{speed}), the UAV's initial and final locations constraints in (\ref{location}), and the GNs' average transmit power constraints in (\ref{10202130}). Consequently, the outage
	probability minimization problem is formulated as
	\begin{align}
		(\mathtt{P2}):~\min\limits_{\{\boldsymbol{q}(t)\},\{P_k(t)\geq0\}}~&O(\{\boldsymbol{q}(t),P_k(t)\})\nonumber\\
		\mathrm{s.t.}~&(\ref{speed}),~(\ref{location}),~\text{and}~(\ref{10202130}).\nonumber
	\end{align}
	It is worth noting that the objective function of problem $(\mathtt{P2})$ is non-convex and even non-smooth due to the indicator function with coupled variables $\boldsymbol{q}(t)$'s and $P_k(t)$'s. In addition, problem $(\mathtt{P2})$ contains an infinite number of optimization variables over continuous time. As a result, problem $(\mathtt{P2})$ is even more challenging to be solved optimally than problem $(\mathtt{P1})$. We will deal with this issue in Section \ref{s5}.

	\section{Proposed Solution to Problem $(\mathtt{P1})$}\label{s4}
	
	In this section, we solve the average data-rate throughput maximization problem $(\mathtt{P1})$ in the delay-tolerant scenario. We first obtain the optimal solution to a relaxed problem of $(\mathtt{P1})$ in the special case with $T\rightarrow\infty$ to gain key engineering insights. Then, based on the optimal solution under the special case, we propose an alternating-optimization-based algorithm to obtain an efficient solution to the original problem $(\mathtt{P1})$ under any finite $T$.
	
	\subsection{Optimal Solution to Relaxed Problem of $(\mathtt{P1})$ with $T\rightarrow\infty$}\label{ss1}
	In this subsection, we consider the special case when the UAV's flight duration $T$ is sufficiently large (i.e., $T\rightarrow\infty$), such that we can ignore the finite flight time of the UAV from one location to another. As a result, the UAV's flight speed constraints in (\ref{speed}) as well as the initial and final locations constraints in (\ref{location}) can be neglected. Therefore, problem $(\mathtt{P1})$ can be relaxed as
	\begin{align}
		(\mathtt{P1.1}):~\max\limits_{\{\boldsymbol{q}(t)\},\{P_k(t)\geq0\}}~&R(\{\boldsymbol{q}(t),P_k(t)\}),~~\mathrm{s.t.}~(\ref{10202130}).\nonumber
	\end{align}
	
	Though problem $(\mathtt{P1.1})$ is still non-convex, it satisfies the so-called time-sharing condition \cite{yu}. Therefore, the strong duality holds between problem $(\mathtt{P1.1})$ and its Lagrange dual problem. As a result, we can optimally solve problem $(\mathtt{P1.1})$ by using the Lagrange duality method \cite{2}. Let $\lambda_k\geq0$ denote the dual variable associated with the $k$-th constraint in (\ref{10202130}). For notational convenience, we define $\boldsymbol\lambda\triangleq[\lambda_1,\ldots,\lambda_K]$. The partial Lagrangian of problem $(\mathtt{P1.1})$ is given as
	\begin{align}
		\mathcal{L}(\{\boldsymbol{q}(t),P_k(t)\},\boldsymbol{\lambda})=&\frac{1}{T}\int_{0}^{T} r(\boldsymbol{q}(t),\{P_k(t)\}){\rm{d}}t \nonumber\\
&-\frac{1}{T}\int_{0}^T\sum_{k=1}^K\lambda_kP_k(t){\rm{d}}t+\sum_{k=1}^K \lambda_k P^{\rm ave}_k.\nonumber
	\end{align}
	The dual function is
	\begin{align}
		g(\boldsymbol{\lambda})&=\max\limits_{\{\boldsymbol{q}(t)\},\{P_k(t)\geq0\}}\mathcal{L}(\{\boldsymbol{q}(t),P_k(t)\},\boldsymbol{\lambda}).\label{201910041530}
	\end{align}
	The dual problem of problem $(\mathtt{P1.1})$ is given by
	\begin{align}
		(\mathtt{D1.1}):\min\limits_{\boldsymbol{\lambda}\succeq \small {\mv 0}}~~&g(\boldsymbol{\lambda}).\nonumber
	\end{align}
	In the following, we solve problem $(\mathtt{P1.1})$ by first obtaining the dual function $g(\boldsymbol{\lambda})$ via solving problem (\ref{201910041530}) and then solving the dual problem ($\mathtt{D1.1}$).

	First, we solve problem (\ref{201910041530}) for finding $g(\boldsymbol{\lambda})$ under given $\boldsymbol{\lambda}$. For notational convenience, let ${\boldsymbol{w}}(t)\!\!=\!\!\left[\!\sqrt{P_{1}(t)},\ldots,\!\sqrt{P_{K}(t)}\!\right]^{\top} \!\!\!\!\in\! \mathbb{R}^{K\times1}$ and $\boldsymbol{h}(\boldsymbol{q}(t))\!\!=\!\!\left[\!\sqrt{\beta_0d_{1}^{-\alpha}(\boldsymbol{q}(t))},\ldots,\!\sqrt{\beta_0d_{K}^{-\alpha}(\boldsymbol{q}(t))}\right]^{\top} \!\!\!\!\in \mathbb{R}^{K\times1}$ denote the GNs' distributed beamforming vector and the combined channel vector at any time instant $t\in \mathcal{T}$, respectively.
	To obtain $g(\boldsymbol{\lambda})$, we decompose problem (\ref{201910041530}) into a set of subproblems, each for one time instant, which are presented in the following with the index $t$ dropped for facilitating the analysis.
	\begin{align}
		\max\limits_{\boldsymbol{q},\boldsymbol{w}\succeq \small {\mv 0}}~\log_2\bigg(1+\frac{\left|{\boldsymbol{w}}^{\top}{\boldsymbol{h}}(\boldsymbol{q})\right|^2}{\sigma^2}\bigg)-\sum_{k=1}^K\lambda_k\|\boldsymbol{e}^{\top}_{k}{\boldsymbol{w}}\|^2,\label{06061626}
	\end{align}
	where $\boldsymbol{e}_{k}\in\mathbb{R}^{K\times1}$ denotes a vector with only the $k$-th element being $1$ and the others being $0$.
	Under any given $\boldsymbol{q}$, problem (\ref{06061626}) is simplified as
	\begin{align}
		\max\limits_{\boldsymbol{w}\succeq \small {\mv 0}}~&\log_2\bigg(1+\frac{\boldsymbol{h}^{\top}\boldsymbol{w}\boldsymbol{w}^{\top}\boldsymbol{h}}{\sigma^2}\bigg)-\text{Tr}(\boldsymbol{B}(\boldsymbol{\lambda})\boldsymbol{w}\boldsymbol{w}^{\top})\label{03011737},
	\end{align}
	where $\text{Tr}(\boldsymbol H)$ refers to the trace of square matrix $\boldsymbol H$, $\boldsymbol B(\boldsymbol{\lambda})\triangleq {\rm \text{Diag}}(\lambda_1,\ldots,\lambda_K)$. In general, we must have $\lambda_k>0$, since otherwise, $g(\boldsymbol{\lambda})$ is not upper bounded. Let $\tilde{\boldsymbol{w}}=\boldsymbol{B}^{1/2}(\boldsymbol{\lambda})\boldsymbol{w}$ and $\tilde{\boldsymbol{h}}=\boldsymbol{B}^{-1/2}(\boldsymbol{\lambda})\boldsymbol{h}$. Then, problem (\ref{03011737}) is recast into
	\begin{align}
		\max\limits_{\tilde{\boldsymbol{w}}\succeq \small {\mv 0}}~&\log_2\bigg(1+\frac{|\tilde{\boldsymbol{h}}^{\top}\tilde{\boldsymbol{w}}|^2}{\sigma^2}\bigg)-\|\tilde{\boldsymbol{w}}\|^2\label{0301178}.
	\end{align}
	Notice that the maximum value of problem (\ref{0301178}) is attained at $\tilde{\boldsymbol{w}}=\sqrt{\tilde{P}}\tilde{\boldsymbol{h}}/\|\tilde{\boldsymbol{h}}\|$ with $\tilde{P}\geq0$. Therefore, problem (\ref{0301178}) can be re-expressed as
	\begin{align}
		\max\limits_{\tilde{P}\geq0}~&\log_2\bigg(1+\frac{\|\tilde{\boldsymbol{h}}\|^2\tilde{P}}{\sigma^2}\bigg)-\tilde{P}\label{030117}.
	\end{align}
	Problem (\ref{030117}) is convex. Hence, by checking the first-derivative of the objective function, we obtain the optimal solution to problem (\ref{030117}) as
	\begin{align}
		 \tilde{P}^{(\boldsymbol\lambda,\boldsymbol{q})}&=\bigg[\frac{1}{\ln2}-\frac{\sigma^2}{\|\tilde{\boldsymbol{h}}(\boldsymbol\lambda,\boldsymbol{q})\|^2}\bigg]^+ \nonumber\\
&=\bigg[\frac{1}{\ln2}-\frac{\sigma^2}{\boldsymbol{h}^{\top}(\boldsymbol{q})\boldsymbol{B}^{-1}(\boldsymbol\lambda)\boldsymbol{h}(\boldsymbol{q})}\bigg]^+,\nonumber
	\end{align}
where $[y]^+\triangleq \max(y,0)$.
	Accordingly, the optimal solution to problem (\ref{03011737}) is given as
	\begin{align}
		 &\boldsymbol{w}^{(\boldsymbol\lambda,\boldsymbol{q})}=\frac{\sqrt{\tilde{P}^{(\boldsymbol\lambda,\boldsymbol{q})}}}{\|\boldsymbol{B}^{-1/2}(\boldsymbol\lambda)\boldsymbol{h}(\boldsymbol{q})\|}\boldsymbol{B}^{-1}(\boldsymbol\lambda)\boldsymbol{h}(\boldsymbol{q}).\nonumber
	\end{align}
	Thus, each GN's optimal power allocation is
	\begin{align}
		 P^{(\boldsymbol\lambda,\boldsymbol{q})}_k=\bigg\|\boldsymbol{e}^{\top}_{k}\frac{\sqrt{\tilde{P}^{(\boldsymbol\lambda,\boldsymbol{q})}}}{\|\boldsymbol{B}^{-1/2}(\boldsymbol\lambda)\boldsymbol{h}(\boldsymbol{q})\|}\boldsymbol{B}^{-1}(\boldsymbol\lambda)\boldsymbol{h}(\boldsymbol{q})\bigg\|^2, \forall k \in \mathcal{K}.\label{power}
	\end{align}
	After substituting $\boldsymbol{w}^{(\boldsymbol\lambda,\boldsymbol{q})}$ into problem (\ref{06061626}), we can obtain the optimal location ${\boldsymbol{q}}^{(\boldsymbol\lambda)}$ for problem (\ref{06061626}) by using the 2D exhaustive search over the region $\mathcal{Z}$, given as
	\begin{align}
		{\boldsymbol{q}}^{(\boldsymbol\lambda)}=&{\rm arg}\max\limits_{\boldsymbol{q}}~\log_2\bigg(1+\frac{\left|{\boldsymbol{w}^{(\boldsymbol\lambda,\boldsymbol{q})}}^{\top}{\boldsymbol{h}}(\boldsymbol{q})\right|^2}{\sigma^2}\bigg) \nonumber\\
&-\sum_{k=1}^K\lambda_k\|\boldsymbol{e}^{\top}_{k}{\boldsymbol{w}^{(\boldsymbol\lambda,\boldsymbol{q})}}\|^2.\label{locationsearch}
	\end{align}
	
	Without loss of generality, suppose that the set of the optimal locations in (\ref{locationsearch}) are given as $\{{\boldsymbol{q}}^{(\boldsymbol\lambda)}_{\nu},\nu\in\mathcal{V}^{(\boldsymbol\lambda)}\triangleq\{1,\ldots,V^{(\boldsymbol\lambda)}\}\}$, with $V^{(\boldsymbol\lambda)}\ge 1$ denoting the number of optimal locations for problem (\ref{locationsearch}). Note that when the optimal solution to problem (\ref{locationsearch}) is non-unique, we can arbitrarily choose any one of ${\boldsymbol q}^{(\boldsymbol\lambda)}_\nu$'s for obtaining $g(\boldsymbol\lambda)$.
	
	Next, we solve the dual problem $(\mathtt{D1.1})$ by minimizing the dual function $g(\boldsymbol\lambda)$. This is implemented via using subgradient-based methods, such as the ellipsoid method \cite{ell}, with the subgradient being $\bigg[P^{\rm ave}_1-P^{(\boldsymbol\lambda,{\boldsymbol{q}}^{(\boldsymbol\lambda)}_{\nu})}_1,\ldots,P^{\rm ave}_K-P^{(\boldsymbol\lambda,{\boldsymbol{q}}^{(\boldsymbol\lambda)}_{\nu})}_K\bigg]$. According to \cite{ell}, the ellipsoid method is convergent and we can finally obtain the optimal solution $\boldsymbol{\lambda}^{\text{opt}}$ to the dual problem $(\mathtt{D1.1})$.
	
	After solving the dual problem $(\mathtt{D1.1})$, it remains to construct the optimal primal solution to problem $(\mathtt{P1.1})$, denoted by $\{{\boldsymbol q}^{\rm opt}(t),P^{\rm opt}_k(t)\}$. In this case, since the optimal solution to problem (\ref{201910041530}) is non-unique in general, we need to time share among these hovering locations to construct the optimal primal solution to problem $(\mathtt{P1.1})$. Let $\tau_\nu$ denote the hovering duration at the optimal location ${\boldsymbol{q}}^{(\boldsymbol\lambda^{\text{opt}})}_{\nu}, \nu\in\mathcal{V}^{(\boldsymbol\lambda^{\text{opt}})}$. In the following, we solve the following problem to obtain the optimal hovering durations for time sharing.
	\begin{align}\label{06201514}
		\max\limits_{\{\tau_\nu\geq0\}}~& \frac{1}{T}\sum\limits_{\nu=1}^{V^{(\boldsymbol\lambda^{\text{opt}})}}\tau_\nu\log_2\bigg(1+\frac{\left|({\boldsymbol{w}}^{(\boldsymbol\lambda^{\text{opt}},{\boldsymbol{q}}^{(\boldsymbol\lambda^{\text{opt}})}_{\nu})})^{\top}{\boldsymbol{h}}({\boldsymbol{q}}^{(\boldsymbol\lambda^{\text{opt}})}_{\nu})\right|^2}{\sigma^2}\bigg) \nonumber\\		 \mathrm{s.t.}~&\frac{1}{T}\sum\limits_{\nu=1}^{V^{(\boldsymbol\lambda^{\text{opt}})}}\tau_\nu\|\boldsymbol{e}^{\top}_{k}{\boldsymbol{w}}^{(\boldsymbol\lambda^{\text{opt}},{\boldsymbol{q}}^{(\boldsymbol\lambda^{\text{opt}})}_{\nu})}\|^2\leq P^{\rm ave}_k,   \forall k\in \mathcal{K}\\
		&\sum\limits_{\nu=1}^{V^{(\boldsymbol\lambda^{\text{opt}})}}\tau_\nu\leq T.\nonumber
	\end{align}
	As problem (\ref{06201514}) is a linear program (LP), the optimal hovering durations $\{\tau^{\rm opt}_\nu\}$ can be obtained by CVX \cite{2}. As a result, problem $(\mathtt{P1.1})$ is optimally solved.
	
	It is observed that the optimal UAV trajectory solution to problem $(\mathtt{P1.1})$ has a multi-location hovering structure, while the GNs' optimal power allocation follows a water-filling-like pattern, dependent on the value of
	$\frac{\sqrt{\tilde{P}^{(\boldsymbol\lambda^{\rm opt},{\boldsymbol{q}}^{(\boldsymbol\lambda^{\text{opt}})}_{\nu})}}\boldsymbol{B}^{-1}(\boldsymbol\lambda^{\rm opt})\boldsymbol{h}(\boldsymbol{q}^{(\lambda^{\text{opt}})}_\nu)}{\|\boldsymbol{B}^{-1/2}(\boldsymbol\lambda^{\rm opt})\boldsymbol{h}(\boldsymbol{q}^{(\lambda^{\text{opt}})}_\nu)\|}$.
	
{\it Complexity Analysis:} The complexity of solving problem $(\mathtt{P1.1})$ includes the complexity of solving dual problem $(\mathtt{D1.1})$ and time sharing problem (\ref{06201514}). In each iteration of the ellipsoid method for optimally solving dual problem $(\mathtt{D1.1})$, the complexity mainly comes from two parts: the $K$ GNs' optimal power allocation in (\ref{power}) and the 2D exhaustive search over the region $\mathcal{Z}$ in (\ref{locationsearch}). The complexity of the first part is $\mathbb{O}(K^2\log K)$ \cite{RanOn2002}, while the complexity of the second part is $\mathbb{O}(1/\varepsilon^2_1)$ \cite{Li2020}, where $\varepsilon_1$ denotes the accuracy of the search. Hence, the complexity of solving dual problem $(\mathtt{D1.1})$ is $\mathbb{O}(K^4\log K/\varepsilon^2_1)$. Furthermore, the complexity of solving linear problem (\ref{06201514}) with $V^{(\boldsymbol\lambda^{\text{opt}})}$ variables and $K+1$ linear constraints is $\mathbb{O}(K^2V^{(\boldsymbol\lambda^{\text{opt}})})$ \cite{2}. Finally, we can conclude that the complexity of solving problem $(\mathtt{P1.1})$ is $\mathbb{O}(K^4\log K/\varepsilon^2_1+K^2V^{(\boldsymbol\lambda^{\text{opt}})})$.
	
	{\it Example 1:}\label{re}
	For obtaining more insights, we consider the special case with two GNs. Without loss of generality, we suppose that the two GNs are deployed at $(-D/2,0,0)$ and $(D/2,0,0)$, where $D$ denotes the distance between the two GNs.
	
	\begin{figure*}[!t]
		\centering
		\subfigure[$D=80$ m]{\includegraphics[width=7cm]{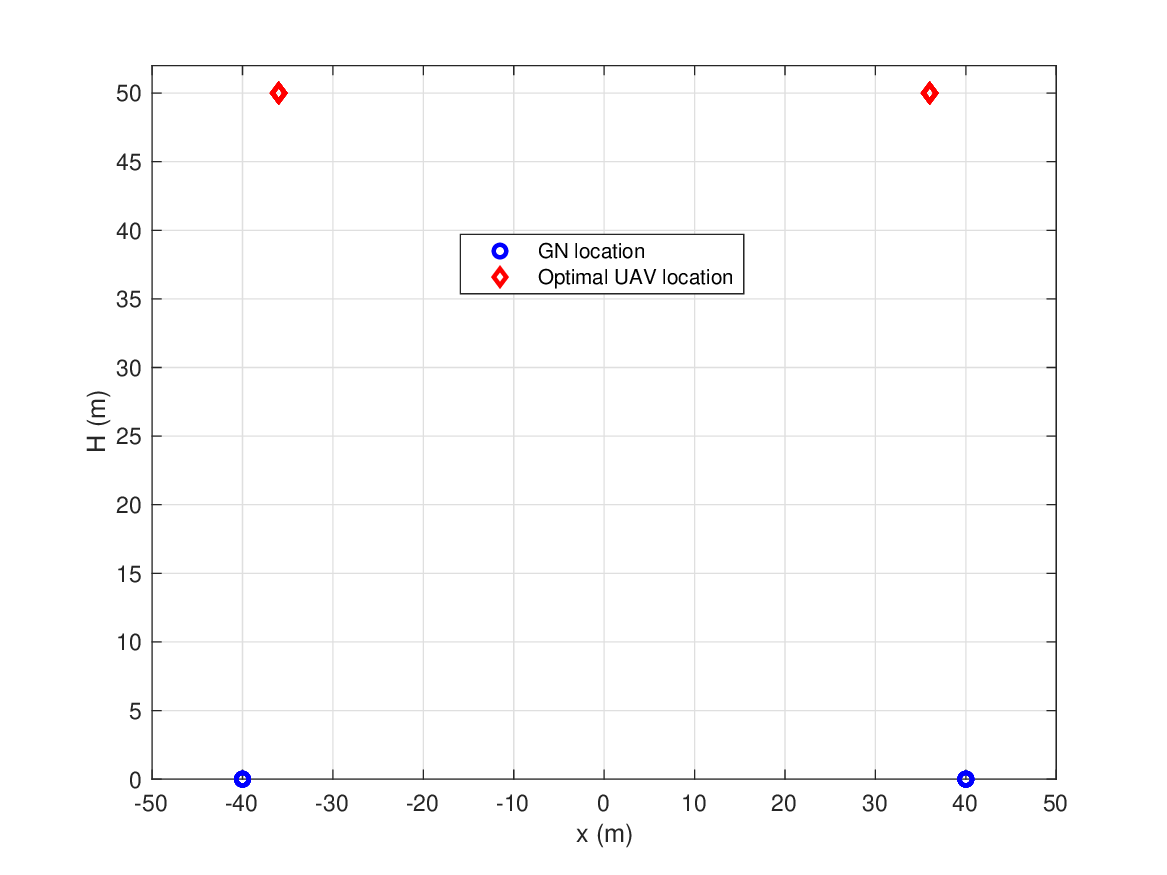}}
		\subfigure[$D=40$ m]{\includegraphics[width=7cm]{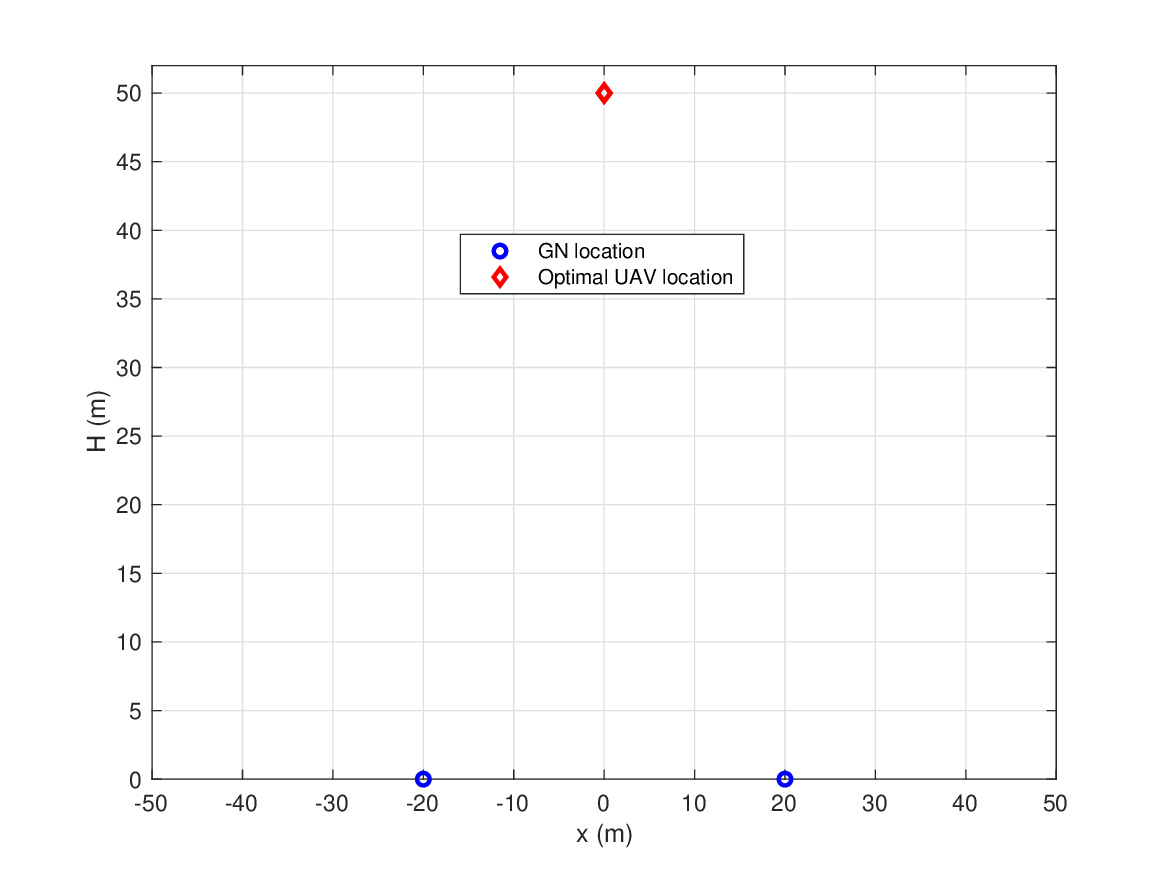}}
		\caption{UAV's optimal hovering locations with different GNs' distances.}
		\label{fig:data_distribution}
	\end{figure*}
	Fig. \ref{fig:data_distribution} shows the UAV's optimal hovering locations with different GNs' distances $D=80$~m in subfigure (a) and $D=40$~m in subfigure (b), where $H=50$~m and $P^{\rm ave}_1=P^{\rm ave}_2=30$~dBm. It is observed that if the two GNs are far away (i.e., $D=80$~m), then the UAV should hover at two symmetric locations with the same hovering time; while if the two GNs are close (i.e., $D=40$~m), the UAV should hover at the middle point of them.
	Table \ref{t1} shows the optimal power and hovering time allocations in Fig. \ref{fig:data_distribution}(a) with $T=10$~s.
	It is observed that the UAV's optimal hovering durations at the two hovering locations are equal due to the symmetric nature of the considered setup; while the GNs' transmit power allocations are different, which have a symmetric structure.
	Moreover, the optimal power allocation of each GN $k$ in Fig. \ref{fig:data_distribution}(b) is obtained at the power $P^{\rm ave}_k$.
	\begin{table}[H]
		\centering
		\caption{GNs' Transmit Power and Hovering Time Allocations}\label{t1}
		\begin{tabular}{|c|c|c|}
			\hline
			Hovering duration&$\tau_1=5$ s&$\tau_2=5$ s\\
			\hline
			GN 1's transmit power&$32.3$~dBm&$25.0$~dBm\\
			\hline
			GN 2's transmit power&$25.0$~dBm&$32.3$~dBm\\
			\hline
		\end{tabular}
	\end{table}
	
	\subsection{Proposed Solution to Problem $(\mathtt{P1})$ with Finite $T$}\label{ss5}
	In this subsection, we consider problem $(\mathtt{P1})$ in the general case with finite $T$. Motivated by the optimal solution to the relaxed problem $(\mathtt{P1.1})$ in the previous subsection, we propose an efficient solution based on the techniques of convex optimization and successive convex approximation (SCA). Towards this end, we first discretize the whole duration $T$ into a finite number of $N$ time slots denoted by the set $\mathcal N\triangleq\{1,...,N\}$, each with equal duration $\delta=T/N$. Let $\boldsymbol{q}[n]$ and $P_k[n]$ denote the UAV's horizontal location and GN $k$'s transmit power at time slot $n$, $k\in \mathcal{K}$, $n\in\mathcal N$. Accordingly, problem $\mathtt{(P1)}$ can be approximated as
	\begin{subequations}
		\begin{align}
			(\mathtt{P1.2}):~&\!\!\!\!\!\max\limits_{\{\boldsymbol{q}[n]\},\{P_k[n]\geq 0\}}\!\!~ \frac{1}{N}\sum\limits_{n=1}^N\log_2(1+{\tt SNR}(\boldsymbol{q}[n],\{P_k[n]\}))\nonumber\\
			\mathrm{s.t.}~&\frac{1}{N}\sum\limits_{n=1}^NP_k[n]\leq P^{\rm ave}_k, \forall k\in \mathcal{K} \label{01160223}\\
			&\|\boldsymbol{q}[n]-\boldsymbol{q}[n-1]\|^2\leq V^2_{\rm max}\delta^2,\forall n\in \mathcal N\label{11142131}\\
			&\boldsymbol{q}[0]=\boldsymbol{q}_{\rm I}, \boldsymbol{q}[N]=\boldsymbol{q}_{\rm F}.\label{11142132}
		\end{align}
	\end{subequations}
	
	Problem $(\mathtt{P1.2})$ is non-convex due to the non-concave objective function. To tackle this issue, we introduce two sets of auxiliary variables $\{a_k[n]\}$ and $\{A[n]\}$, $k\in\mathcal K,n\in\mathcal N$. We assume that $a_k[n]$ and $A[n]$ are the lower bounds of $\sqrt{\frac{P_k[n]\beta_0}{(\|\boldsymbol{q}[n]-\boldsymbol{s}_k\|^2+H^2)^{\alpha/2}}}$ and $\left(\sum_{k=1}^Ka_k[n]\right)^2$, respectively. Problem $(\mathtt{P1.2})$ is re-expressed as
	\begin{subequations}
		\begin{align}
			(\mathtt{P1.3})&:\!\!\!\!\!\!\max\limits_{\{\boldsymbol{q}[n]\},\{P_k[n]\geq0\},\{A[n]\},\{a_k[n]\}}\!\!\!~ \frac{1}{N}\!\!\sum\limits_{n=1}^N\log_2\!\left(1+A[n]/\sigma^2\right)\nonumber\\
			\mathrm{s.t.}~& A[n]\leq\left(\sum_{k=1}^Ka_k[n]\right)^2, \forall n\in \mathcal{N}\label{123011022}\\
			&\!\!\!\!\!\!\!\!\!\!\!\!\!\! a_k[n]\leq\sqrt{\frac{P_k[n]\beta_0}{(\|\boldsymbol{q}[n]-\boldsymbol{s}_k\|^2+H^2)^{\alpha/2}}}, \forall k\in \mathcal{K}, n\in \mathcal{N}\label{123011032}\\
            &(\ref{01160223}),~(\ref{11142131}),~\text{and}~(\ref{11142132}).\nonumber
		\end{align}
	\end{subequations}
	Note that at the optimal solution to problem (P1.3), all the constraints in (\ref{123011022}) and (\ref{123011032}) must be met with strict equality, since otherwise, we can increase $a_k[n]$ and $A[n]$, thus further increasing the objective value of problem (P1.3). Problem $(\mathtt{P1.3})$ is still non-convex due to the non-convex constraints in (\ref{123011022}) and (\ref{123011032}).
	Next, we solve problem $(\mathtt{P1.3})$ by alternatively optimizing the UAV's trajectory and the GNs' power allocation.
	\subsubsection{Trajectory Optimization}
	Under given GNs' power allocation $\{P_k[n]\geq 0\}$, we optimize the UAV's trajectory $\{\boldsymbol{q}[n]\}$ with variables $\{A[n]\}$ and $\{a_k[n]\}$ for problem $(\mathtt{P1.3})$ by adopting the SCA technique. To deal with the non-convex constraints in (\ref{123011022}) and (\ref{123011032}), we update the UAV's trajectory $\{\boldsymbol{q}[n]\}$ and $\{a_k[n]\}$ in an iterative manner by approximating the non-convex problem into a convex problem. Let $\{\boldsymbol{q}^{(i)}[n]\}$ and $\{a_k^{(i)}[n]\}$ denote the local points at the $i$-th iteration. Under given UAV's trajectory $\{\boldsymbol{q}^{(i)}[n]\}$ and $\{{a}^{(i)}_k[n]\}$, since any convex function is globally lower-bounded by its first-order Taylor expansion at any point, we have the lower bounds for $\sqrt{\frac{P_k[n]\beta_0}{(\|\boldsymbol{q}[n]-\boldsymbol{s}_k\|^2+H^2)^{\alpha/2}}}$ and $\left(\sum_{k=1}^Ka_k[n]\right)^2$ as follows.
	\begin{align}
		 &\sqrt{\!\frac{P_k[n]\beta_0}{\!\!(\|\boldsymbol{q}[n]\!-\!\boldsymbol{s}_k\|^2\!\!+\!\!H^2\!)^{\alpha/2}}}\!\!\geq\!\!\sqrt{P_k\beta_0}\bigg(\!\!(\|\boldsymbol{q}^{(i)}[n]\!-\!\boldsymbol{s}_k\|^2\!\!+\!\!H^2)^{-\alpha/4}\nonumber\\
		 &~-\frac{\alpha(\|\boldsymbol{q}[n]-\boldsymbol{s}_k\|^2-\|\boldsymbol{q}^{(i)}[n]-\boldsymbol{s}_k\|^2)}{4(\|\boldsymbol{q}^{(i)}[n]-\boldsymbol{s}_k\|^2+H^2)^{\alpha/4+1}}\!\!\bigg)\!\triangleq\!{a^{{\rm low}}_{k(i)}}(\boldsymbol{q}[n]),\label{06191622}
	\end{align}
	\begin{align}
		&\bigg(\sum_{k=1}^Ka_k[n]\bigg)^2\geq\bigg(\sum_{k=1}^Ka^{(i)}_k[n]\bigg)^2+2\bigg(\sum_{k=1}^Ka^{(i)}_k[n]\bigg) \nonumber\\
&\times \bigg(\sum_{k=1}^Ka_k[n]-\sum_{k=1}^Ka^{(i)}_k[n]\!\!\bigg)\triangleq A^{{\rm low}}_{(i)}(a_k[n]).\label{06191623}
	\end{align}
	In each iteration $i$ with given local points $\{\boldsymbol{q}^{(i)}[n]\}$ and $\{{a}^{(i)}_k[n]\}$, we replace $\sqrt{\frac{P_k[n]\beta_0}{(\|\boldsymbol{q}[n]-\boldsymbol{s}_k\|^2+H^2)^{\alpha/2}}}$ and $\bigg(\sum_{k=1}^Ka_k[n]\bigg)^2$ as their lower bounds ${a^{{\rm low}}_{k(i)}}(\boldsymbol{q}[n])$ and $A^{{\rm low}}_{(i)}(a_k[n])$, respectively. As a result, the trajectory optimization problem is changed to a convex optimization problem, which is denoted as problem $(\mathtt{P1.3.1})$ and can be optimally solved by CVX \cite{2}.
	
	\subsubsection{Power Allocation}\label{Power Allocation}
	Under any given UAV trajectory $\{\boldsymbol{q}[n]\}$, we optimize the GNs' power allocation $\{P_k[n]\geq 0\}$ together with $\{A[n]\}$ and $\{a_k[n]\}$ for problem $(\mathtt{P1.3})$ by using the SCA technique as well. In this case, only the constraints in (\ref{123011022}) are non-convex. Similarly as for optimizing the UAV trajectory, we replace $\bigg(\sum_{k=1}^Ka_k[n]\bigg)^2$ in (\ref{123011022}) as its lower bound $A^{{\rm low}}_{(i)}(a_k[n])$ in (\ref{06191623}) to approximate the non-convex terms into convex forms and denote the approximate problem as problem $(\mathtt{P1.3.2})$, so as to optimize the GNs' power allocation iteratively, for which the details are omitted for brevity.
	
	According to \cite{WuJoint2018}, by alternately optimizing the UAV trajectory and GNs' power allocation, we can obtain a converged solution to problem $\mathtt{(P1.3)}$, thus efficiently solving problem $\mathtt{(P1.2)}$.
	
{\it Complexity Analysis:} The complexity of the proposed alternating optimization-based approach for solving problem $\mathtt{(P1.3)}$ includes the complexities of solving the power allocation sub-problem and trajectory design sub-problem. By using SCA, these two sub-problems are solved in an iterative manner by respectively approximating as two convex optimization problems: sub-problem $\mathtt{(P1.3.1)}$ and sub-problem $\mathtt{(P1.3.2)}$. According to \cite{2} and \cite{ValiulahiMulti2020}, CVX uses interior point method with Newton steps to solve the convex optimization problem, whose complexity is $\mathbb{O}((B+C)^{1.5}B^2)$, where $B$ is the dimension of optimization variables and $C$ is the number of constraints. For problem $\mathtt{(P1.3.1)}$, we have $B=(K+2)N$ and $C=(K+2)N+2$, while for problem $\mathtt{(P1.3.2)}$, we have $B=2KN+N$ and $C=KN+N+K$. Thus, the complexities of solving the two sub-problems are both $\mathbb{O}(K^{3.5}N^{3.5})$. Then, the computation complexity for solving problem $\mathtt{(P1.3)}$ is $\mathbb{O}(L_{\rm alt}(L_{\rm SCA1}+L_{\rm SCA2})K^{3.5}N^{3.5})$, where $L_{\rm alt}$, $L_{\rm SCA1}$, and $L_{\rm SCA2}$ are the iteration numbers of alternating optimization for problem $\mathtt{(P1.3)}$, SCA for sub-problem $\mathtt{(P1.3.1)}$, and SCA for sub-problem $\mathtt{(P1.3.2)}$, respectively.

	\begin{remark}\label{initial_trajectory}
		It is worth noting that the performance of the alternating optimization-based approach critically depends on the initial point chosen for iteration. In this paper, we consider the following three trajectory designs as the potential initial point.
		\begin{itemize}
			\item \textbf{Fly-hover-fly trajectory with power design}:
			The UAV first flies straightly from the initial location to one optimized fixed location $(x^{\rm fix},y^{\rm fix},H)$, and hovers at this location as long as possible, and finally flies to the final location at the maximum flight speed.
			The fixed location $(x^{\rm fix},y^{\rm fix},H)$ is obtained by using a 2D exhaustive search over the region in $\mathcal{Z}$ during the mission time to maximize the received SNR at the UAV, during which each GN $k$ employs the fixed power $P^{\rm ave}_k$.
			Thus, the flying time $T^{\rm FHF}_{\rm fly}$ is $\|\boldsymbol q_I-(x^{\rm fix},y^{\rm fix})\|/V_{\rm max}+\|(x^{\rm fix},y^{\rm fix})-\boldsymbol q_F\|/V_{\rm max}$ and the hovering duration at the optimized location is given as $T^{\rm FHF}_{\rm hov}=T-T^{\rm FHF}_{\rm fly}$.
			Under such a trajectory, the GNs' power allocation can be obtained by solving the power allocation problem in $\mathtt{(P1.3)}$.
			\item \textbf{Successive hover-and-fly trajectory with power design}: The UAV flies from the initial location to successively visit the optimized hovering locations to problem $\mathtt{(P1.1)}$, then hovers at these locations, and finally flies to the final location at the maximum flight speed. During the flight, we choose the minimum flying path by solving the traveling salesman problem (TSP) \cite{XieThroughput2019}. Then, we have the minimum flying time $T^{\rm SHF}_{\rm fly}$ and the hovering duration at each optimized location can be obtained similarly by solving problem $\mathtt{(P1.1)}$, with total hovering time given by $T^{\rm SHF}_{\rm hov}=T-T^{\rm SHF}_{\rm fly}$. Under such a trajectory, the GNs' power allocation can be obtained by solving the power allocation problem in problem $\mathtt{(P1.3)}$.
			\item \textbf{Power design only}: The UAV flies from the initial location to the final location directly with a constant flight speed $\|\boldsymbol q_I-\boldsymbol q_F\|/T$. Under such a trajectory, the GNs' power allocation can be obtained by solving the power allocation problem in $\mathtt{(P1.3)}$.
		\end{itemize}
		
		Note that the minimum flying time in each trajectory design should be no larger than the UAV flight duration $T$ to guarantee a feasible trajectory.
		In this case, under any given $T$, we choose the one which has the best performance as the initial point of our proposed SCA-based algorithm.
	\end{remark}
	
	\section{Proposed Solution to Problem $(\mathtt{P2})$}\label{s5}	
	In this section, we address the outage probability minimization problem $(\mathtt{P2})$ in the delay-sensitive application scenario. We first obtain the optimal solution to a relaxed problem of $(\mathtt{P2})$ in the special case with $T\rightarrow\infty$ to gain key engineering insights. Then, based on the optimal solution under the special case, we propose an alternating-optimization-based algorithm to obtain an efficient solution to the original problem $(\mathtt{P2})$ under any finite $T$.
	
	\subsection{Optimal Solution to Relaxed Problem of $(\mathtt{P2})$ with $T\rightarrow\infty$}\label{ss6}
	In this subsection, we consider the special case that the UAV's flight duration $T$ is sufficiently large (i.e., $T\rightarrow\infty$). Similarly as for problem $(\mathtt{P1})$, problem $(\mathtt{P2})$ can be relaxed as
	\begin{align}
		(\mathtt{P2.1}):~\min\limits_{\{\boldsymbol{q}(t)\},\{P_k(t)\geq0\}}~&O(\{\boldsymbol{q}(t),P_k(t)\}),~~\mathrm{s.t.}~(\ref{10202130}).\nonumber
	\end{align}
	Though problem $(\mathtt{P2.1})$ is non-convex, it satisfies the so-called time-sharing condition \cite{yu}. Therefore, the strong duality holds between problem $(\mathtt{P2.1})$ and its Lagrange dual problem. As a result, we can optimally solve problem $(\mathtt{P2.1})$ by using the Lagrange duality method \cite{2} as follows. Let $\mu_k\geq0$ denote the dual variable associated with the $k$-th constraint in (\ref{10202130}). For notational convenience, we define $\boldsymbol\mu\triangleq[\mu_1,\ldots,\mu_K]$.
	The partial Lagrangian of problem $(\mathtt{P2.1})$ is given as
	\begin{align}
&\tilde{\mathcal{L}}(\{\boldsymbol{q}(t)\},\{P_k(t)\},\boldsymbol\mu)=
		\frac{1}{T}\int_{0}^T\mathbbm{1}\left({\tt SNR}(\boldsymbol{q}(t),\{P_k(t)\})\right){\rm{d}}t \nonumber\\
&~~~~~~+ \frac{1}{T}\int_{0}^T\sum_{k=1}^K\mu_kP_k(t){\rm{d}}t-\sum_{k=1}^K\mu_kP_k^{\rm ave}.\nonumber
	\end{align}
	The dual function is
	\begin{align}
		&\tilde{g}(\boldsymbol\mu)=\min\limits_{\{\boldsymbol{q}(t)\},\{P_k(t)\ge 0\}}\tilde{\mathcal{L}}(\{\boldsymbol{q}(t)\},\{P_k(t)\},\boldsymbol\mu).\label{03010820111}
	\end{align}
	The dual problem of problem $(\mathtt{P2.1})$ is given by
	\begin{align}
		(\mathtt{D2.1}):\max\limits_{\boldsymbol\mu\succeq \small {\mv 0}}~~&\tilde{g}(\boldsymbol\mu).\nonumber
	\end{align}
	In the following, we solve problem $(\mathtt{P2.1})$ by first obtaining the dual function $\tilde{g}(\boldsymbol\mu)$ and then solving the dual problem ($\mathtt{D2.1}$). First, to obtain $\tilde{g}(\boldsymbol\mu)$, we solve problem (\ref{03010820111}) by solving a set of subproblems, each for a time instant in the following, in which the index $t$ is dropped for facilitating the analysis.
	\begin{align}
		\min\limits_{\boldsymbol{q},\{P_k\geq0\}}~\mathbbm{1}({\tt SNR}(\boldsymbol{q},\{P_k\}))+\sum_{k=1}^K\mu_kP_k.\label{06061630}
	\end{align}
	To solve problem (\ref{06061630}), we consider the following two cases when $\mathbbm{1}\left({\tt SNR}(\boldsymbol{q},\{P_k\})\right)$ equals one and zero, respectively.
	
	\subsubsection{Outage case}\label{outage} First, consider that $\mathbbm{1}\left({\tt SNR}(\boldsymbol{q},\{P_k\})\right)=1$. In this case, the outage occurs, and thus GN's optimal power allocation is $\{P^{(\boldsymbol\mu,\boldsymbol{q})}_k=0\}$, and UAV's optimal location $\boldsymbol{q}^{(\boldsymbol\mu)}$ can be any arbitrary value in $\mathcal{Z}$. Accordingly, the optimal value for problem (\ref{06061630}) is $1$.
	
	\subsubsection{Non-outage case}\label{non-outage} Next, consider that $\mathbbm{1}\!\left({\tt SNR}(\boldsymbol{q},\{P_k\})\right)\!\!=\!\!\!0$. In this case, we solve problem (\ref{06061630}) by first deriving the GNs' power allocation under any given UAV's location $\boldsymbol{q}$ and then searching over $\boldsymbol{q}$ via a 2D exhaustive search over $\mathcal{Z}$. Under given $\boldsymbol{q}$ and defining $\rho_k=\sqrt{P_k}, \forall k\in\mathcal{K}$, problem (\ref{06061630}) is reduced as
	\begin{align}
		\min\limits_{\{\rho_k\geq 0\}}~&\sum_{k=1}^K\mu_k\rho^2_k\label{06061611111}\\
		\mathrm{s.t.}~&\sum_{k=1}^{K}\rho_k\sqrt{\beta_0d_{k}^{-\alpha}(\boldsymbol{q})}\geq\sqrt{\gamma_{\rm min}}\sigma.\nonumber
	\end{align}
	If $\mu_k>0, \forall k\in \mathcal{K}$, then problem (\ref{06061611111}) is a convex problem. By checking the Karush-Kuhn-Tucker (KKT) conditions, we have the optimal solution as
	\begin{align}
		\rho^{(\boldsymbol\mu,\boldsymbol{q})}_k=\frac{\sqrt{\gamma_{\rm min}\beta_0d_{k}^{-\alpha}(\boldsymbol{q})}\sigma}{\bigg(\sum_{k=1}^K
			(\beta_0d_{k}^{-\alpha}(\boldsymbol{q})/\mu_k)\bigg)\mu_k}.\label{power2}
	\end{align}
	If there exists any $k\in \mathcal{K}$ such that $\mu_k=0$, then the optimal value of problem (\ref{06061611111}) is zero, which is attained by setting $\rho^{(\boldsymbol\mu,\boldsymbol{q})}_k$ to be sufficiently large and $\rho^{(\boldsymbol\mu,\boldsymbol{q})}_{\bar k}=0, \forall \bar k\neq k$. Therefore, we obtain $P^{(\boldsymbol\mu,\boldsymbol{q})}_k={\rho^{(\boldsymbol\mu,\boldsymbol{q})}_k}^2$. By substituting $P^{(\boldsymbol\mu,\boldsymbol{q})}_k$ into problem (\ref{06061630}), we obtain the optimal UAV location $\boldsymbol{q}^{(\boldsymbol\mu)}$ by using the 2D exhaustive search over $\mathcal{Z}$, given as
	\begin{align}
		\boldsymbol{q}^{(\boldsymbol\mu)}=&\arg \min_{\boldsymbol{q}}\mathbbm{1}\left({\tt SNR}(\boldsymbol{q},\{P^{(\boldsymbol\mu,\boldsymbol{q})}_k\})\right)+\sum_{k=1}^K\mu_kP^{(\boldsymbol\mu,\boldsymbol{q})}_k.\label{Searchq}
	\end{align}
	Accordingly, the obtained power allocation is given by $\{P^{(\boldsymbol\mu,\boldsymbol{q}^{(\boldsymbol\mu)})}_{k}\}$ and the optimal value for problem (\ref{06061630}) is $\sum_{k=1}^K\mu_kP^{(\boldsymbol\mu,\boldsymbol{q}^{(\boldsymbol\mu)})}_k$.
	Without loss of generality, suppose that the set of the optimal locations is $\{{\boldsymbol{q}}^{(\boldsymbol\mu)}_{\tilde{\nu}}, \tilde{\nu}\in\tilde{\mathcal{V}}^{(\boldsymbol\mu)}\triangleq\{1,\ldots,\tilde{V}^{(\boldsymbol\mu)}\}\}$, with $\tilde{V}^{(\boldsymbol\mu)}\ge 1$ denoting the number of optimal locations for problem (\ref{Searchq}).
	Note that when the optimal location for problem (\ref{Searchq}) is non-unique, we can arbitrarily choose any one of ${\boldsymbol q}^{(\boldsymbol{\mu})}_{\tilde{\nu}}$'s for obtaining $\tilde{g}(\boldsymbol{\mu})$.
	
	By comparing the corresponding optimal values under $\mathbbm{1}\left({\tt SNR}(\boldsymbol{q},\{P_k\})\right)=1$ and $\mathbbm{1}\left({\tt SNR}(\boldsymbol{q},\{P_k\})\right)=0$, we can obtain the optimal solution to problem (\ref{06061630}) as the one achieving the smaller optimal value. Therefore, the dual function $\tilde{g}(\boldsymbol\mu)$ is obtained.
	
	Next, we solve the dual problem $(\mathtt{D2.1})$ by maximizing the dual function $\tilde{g}(\boldsymbol\mu)$. This is implemented via using subgradient-based methods, such as the ellipsoid method \cite{ell}, with the subgradient being $[P^{(\boldsymbol\mu,{\boldsymbol{q}}^{(\boldsymbol\mu)}_{\tilde{\nu}})}_1-P^{\rm ave}_1,\ldots,P^{(\boldsymbol\mu,{\boldsymbol{q}}^{(\boldsymbol\mu)}_{\tilde{\nu}})}_K-P^{\rm ave}_K]$. We denote the optimal dual solution to the dual problem $(\mathtt{D2.1})$ as $\boldsymbol{\mu}^{\text{opt}}$.
	
	At the optimal dual solution $\boldsymbol{\mu}^{\text{opt}}$, we need to deal with the following two cases.
	\begin{itemize}
		
		\item
		If $\sum_{k=1}^K\mu^{\text{opt}}_kP^{({\boldsymbol\mu}^{\text{opt}},\boldsymbol{q}^{({\boldsymbol\mu}^{\text{opt}})})}_k=1$, (i.e., the outage case {\it 1)} and the non-outage case {\it 2)} lead to the same optimal value of problem (\ref{06061630})), then we need to time share between case {\it 1)} and case {\it 2)} to construct the primal optimal trajectory and power allocation, denoted by $\{\tilde{\boldsymbol{q}}^{\rm opt}(t)\}$ and $\{\tilde{P}_k^{\rm opt}(t)\}$, respectively. Notice that under $\boldsymbol{\mu}^{\text{opt}}$, the optimal solution to problem (\ref{06061630}) is generally non-unique in case {\it 2)}. Therefore, we also need to time share among these UAV locations and the corresponding power allocation strategies to construct the primal optimal solution to problem $\mathtt{(P2.1)}$. Let $\tilde{\tau}_{\tilde{\nu}}$ denote the UAV's hovering duration at the location $\boldsymbol{q}^{(\boldsymbol\mu^{\text{opt}})}_{\tilde{\nu}}$, $\tilde{\nu} \in \tilde{\mathcal{V}}^{(\boldsymbol\mu^{\text{opt}})}$. In the following, we solve the following problem to obtain the optimal hovering durations for time sharing.
		\begin{align}
			\min\limits_{\{\tilde{\tau}_{\tilde{\nu}}\geq0\}}~& \frac{1}{T}\bigg(T-\sum\limits_{\tilde{\nu}=1}^{\tilde{V}^{(\boldsymbol\mu^{\text{opt}})}}\tilde{\tau}_{\tilde{\nu}}\bigg)\label{06201515}\\
			 \mathrm{s.t.}~&\frac{1}{T}\sum\limits_{\tilde{\nu}=1}^{\tilde{V}^{(\boldsymbol\mu^{\text{opt}})}}\tilde{\tau}_{\tilde{\nu}}P^{(\boldsymbol\mu^{\text{opt}},\boldsymbol{q}^{(\boldsymbol\mu^{\text{opt}})}_{\tilde{\nu}})}_{k}\leq P_k^{\rm ave},   \forall k\in \mathcal{K}\nonumber\\
			&\sum\limits_{\tilde{\nu}=1}^{\tilde{V}^{(\boldsymbol\mu^{\text{opt}})}}\tilde{\tau}_{\tilde{\nu}}\leq T.\nonumber
		\end{align}
		
		In problem (\ref{06201515}), we have omitted the time duration when outage occurs, which should be $T-\sum\limits_{\tilde{\nu}=1}^{\tilde{V}^{(\boldsymbol\mu^{\text{opt}})}}\tilde{\tau}_{\tilde{\nu}}$. As problem (\ref{06201515}) is a linear program, the optimal hovering durations $\{\tilde{\tau}^{\text{opt}}_{\tilde{\nu}}\}$ can be obtained by CVX \cite{2}. Therefore, problem ($\mathtt{P2.1}$) is finally optimally solved.
		
		Note that at the optimal solution, the UAV hovers at multiple locations $\{\boldsymbol{q}^{(\boldsymbol\mu^{\text{opt}})}_{\tilde{\nu}}\}$ each with duration $\tilde{\tau}^{\text{opt}}_{\tilde{\nu}}$ to collect data from GNs, and the GNs adopt an on-off power allocation, i.e., the GNs are active to send messages with properly designed power allocation (i.e., $P^{({\boldsymbol\mu}^{\text{opt}},\boldsymbol{q}^{({\boldsymbol\mu}^{\text{opt}})})}_k$) when no outage occurs, but inactive with zero transmit power when outage occurs. Also note that the duration with outage occurring is given by $\tilde{\tau}^{\text{opt}}_0=T-\sum\limits_{\tilde{\nu}=1}^{\tilde{V}^{(\boldsymbol\mu^{\text{opt}})}}\tilde{\tau}_{\tilde{\nu}}$, with the resulting outage probability being $\tilde{\tau}^{\text{opt}}_0/T$.
		
		\item
		If $\sum_{k=1}^K\mu^{\text{opt}}_kP^{({\boldsymbol\mu}^{\text{opt}},\boldsymbol{q}^{({\boldsymbol\mu}^{\text{opt}})})}_k<1$ (i.e., non-outage occurs), then the UAV can achieve non-outage communication over the whole mission period. However, in this case it becomes difficult to directly find the feasible or optimal solution to problem $\mathtt{(P2.1)}$. Hence, we use an additional step to obtain the primal optimal solution to problem $\mathtt{(P2.1)}$.
		In this case, we reduce the transmit power at all GNs by reducing $P^{\rm ave}_k$ as $\tilde{\alpha}P^{\rm ave}_k$, with $0<\tilde{\alpha}<1$. We solve problem $\mathtt{(P2.1)}$ under different $\tilde{\alpha}$ together with a bisection over $\tilde{\alpha}$, in order to find a $\tilde{\alpha}^*$ such that the resultant outage probability is slightly lager than $0$. In this case, the obtained $\{\tilde{\boldsymbol{q}}^{\rm opt}(t)\}$ can be used as a feasible solution to problem $\mathtt{(P2.1)}$. Accordingly, by finding the feasible power allocations at these locations, an optimal solution of $\{\tilde{\boldsymbol{q}}^{\rm opt}(t),\tilde{P}_k^{\rm opt}(t)\}$ to problem $\mathtt{(P2.1)}$ can finally be obtained.
	\end{itemize}
	
	{\it Complexity Analysis:} The complexity of solving problem $(\mathtt{P2.1})$ includes solving dual problem $(\mathtt{D2.1})$ and time sharing problem (\ref{06201515}). Similar to problem $(\mathtt{P1.1})$, we can obtain that the complexities of solving dual problem $(\mathtt{D2.1})$ and linear problem (\ref{06201515}) are $\mathbb{O}(K^3/\varepsilon^2_2)$ and  $\mathbb{O}(K^2\tilde{V}^{(\boldsymbol\mu^{\text{opt}})})$ \cite{2}, where $\varepsilon_2$ denotes the accuracy of the search for problem $(\mathtt{D2.1})$, $\tilde{V}^{(\boldsymbol\mu^{\text{opt}})}$ and $K+1$ are the numbers of variables and linear constraints for problem (\ref{06201515}), respectively.
It is noticed that, when no outage occurs, we need to add one more outer step (i.e., bisection searching over $P^{\rm ave}_k$) to solve problem $(\mathtt{P2.1})$, whose complexity is $\mathbb{O}(\log(1/\varepsilon_3))$, in which $\varepsilon_3$ denotes the accuracy of the bisection search. Therefore, the complexity of solving problem $(\mathtt{P2.1})$ is $\mathbb{O}((\log(1/\varepsilon_3))(K^3/\varepsilon^2_2+K^2\tilde{V}^{(\boldsymbol\mu^{\text{opt}})}))$.
	
\begin{figure*}[!htb]
		\centering
		\subfigure[$D=80$ m]{\includegraphics[width=7cm]{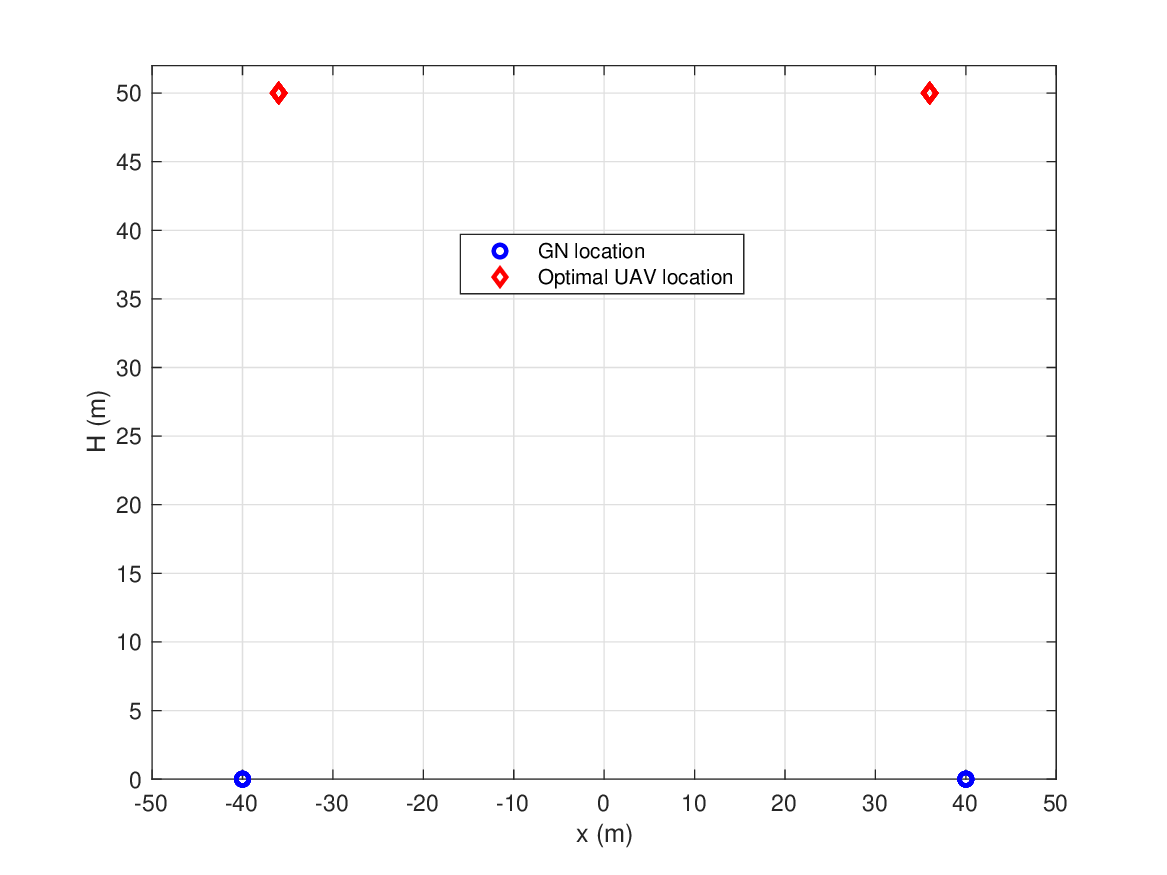}}
		\subfigure[$D=40$ m]{\includegraphics[width=7cm]{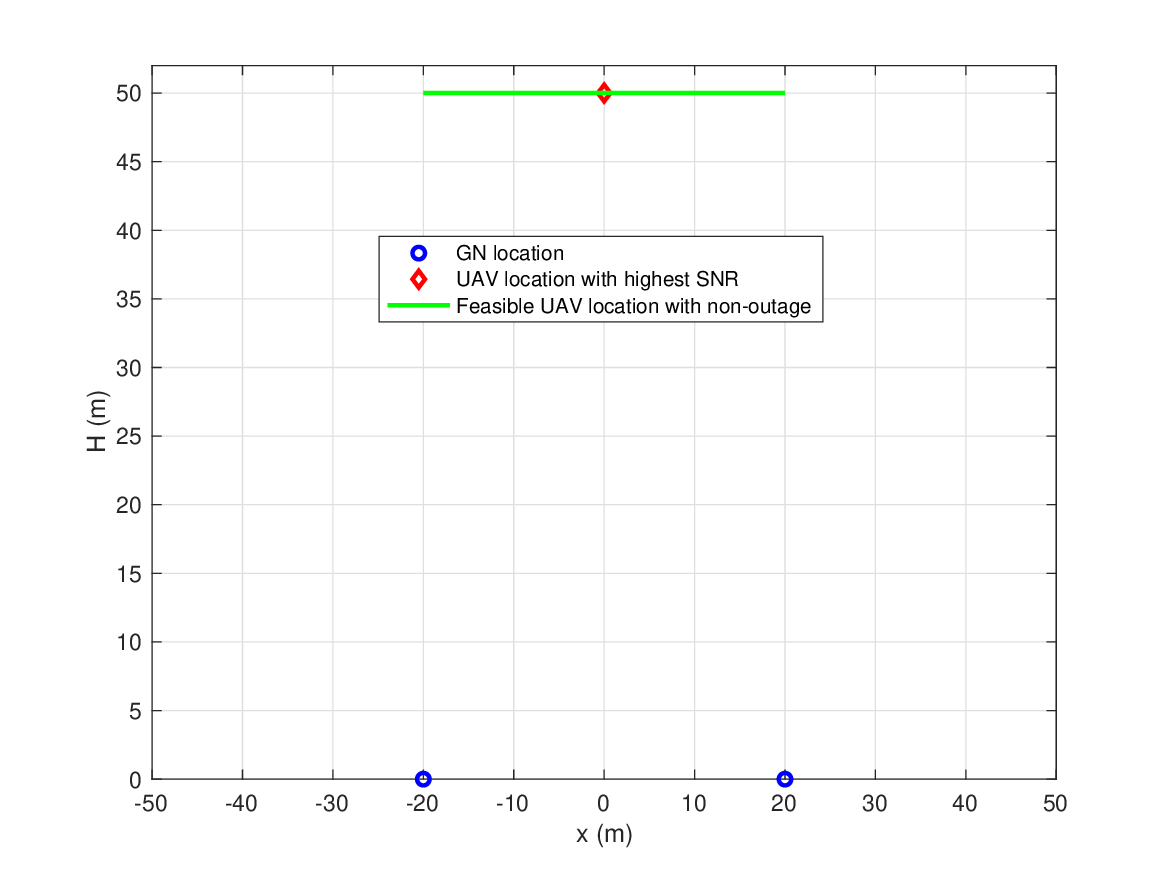}}
		\caption{UAV's optimal hovering locations with different GNs' distances.}
		\label{fig:data_distribution1}
	\end{figure*}

	{\it Example 2:}
	For obtaining more insights, we consider the special case with two GNs, where the setup is the same as {\it Example 1}. Besides, we set $\gamma_{\rm min}=17$ dB.
	Fig. \ref{fig:data_distribution1}(a) and Fig. \ref{fig:data_distribution1}(b) show the optimal hovering locations with different GNs' distances being $D=80$~m and $D=40$~m, respectively. When $D = 80$ m, the optimal hovering locations are observed to be the same as those in Fig. \ref{fig:data_distribution}(a) in {\it Example 1}; while when $D = 40$ m, the UAV can hover at any point within the desirable flight region to achieve non-outage communication (i.e., $O(\{\tilde{\boldsymbol{q}}^{\rm opt}(t),\tilde{P}_k^{\rm opt}(t)\})=0$). This is due to the fact that when the GNs are close and have sufficient transmit power, they can easily meet the minimum SNR requirement at the UAV when the UAV is within the indicated region, as shown in Fig. \ref{fig:data_distribution1}(b). Notice that in Fig. \ref{fig:data_distribution1}(b) we also show the hovering location that leads to the highest SNR, which is observed to be exactly the optimal hovering location in Fig. \ref{fig:data_distribution}(b) in {\it Example 1}.
	
Table \ref{t2} shows the optimal power and hovering time durations in Fig. \ref{fig:data_distribution1}(a) with $T=10$~s.
	It is observed that the system is non-outage for $8.24$~s and outage for $1.76$~s. When the system is non-outage, similar observation is shown as {\it Example 1} and the optimal trajectory has the similar multi-location-hovering structure as in {\it Example 1}.
	Noting that, though the UAV's optimal hovering locations are similar as in {\it Example 1}, the GNs' power allocation in {\it Example 2} is different. In particular, the GNs need to focus more power at each optimal hovering location to satisfy the SNR requirement. Therefore, the GNs in {\it Example 2} adopt an on-off power allocation and give up the transmission at some time, while the GNs in {\it Example 1} transmit continuously based on a water-filling-like power allocation to balance the data-rate throughput over time. Moreover, at the non-outage time in Fig. \ref{fig:data_distribution1}(a), the ratio between the two GNs' transmit powers at each optimized hovering location is the same as that in Fig. \ref{fig:data_distribution}(a) in {\it Example 1}; while in Fig. \ref{fig:data_distribution1}(b), the ratio between the two GNs' transmit powers at each optimized hovering location is non-unique in general. Here, we set the power allocation the same as that in Fig. \ref{fig:data_distribution}(b) in {\it Example 1}, since such a power allocation leads to the maximized SNR in both scenarios.
	
	\begin{table}[H]
		\centering
		\caption{GNs' Transmit Power and Hovering Time Allocations}\label{t2}
		\begin{tabular}{|c|c|c|c|}
			\hline
			Hovering duration&$\tilde{\tau}_0=1.76$ s&$\tilde{\tau}_1=4.12$ s&$\tilde{\tau}_2=4.12$ s\\
			\hline
			GN 1's transmit power&0&$33.1$~dBm&$25.8$~dBm\\
			\hline
			GN 2's transmit power&0&$25.8$~dBm&$33.1$~dBm\\
			\hline
		\end{tabular}
	\end{table}

	\subsection{Proposed Solution to Problem $(\mathtt{P2})$ with Finite $T$}\label{s10}	
	In this subsection, we consider problem $(\mathtt{P2})$ in the general case with finite $T$. Motivated by the optimal solution to the relaxed problem $(\mathtt{P2.1})$ in the previous subsection, we propose an efficient solution based on the techniques of convex optimization and SCA. Towards this end, we first discretize the whole duration $T$ into a finite number of $\tilde{N}$ time slots denoted by the set $\tilde{\mathcal N}\triangleq\{1,...,\tilde{N}\}$, each with equal duration $\tilde{\delta}=T/\tilde{N}$. Accordingly, problem $\mathtt{(P2)}$ is re-expressed as
	\begin{subequations}
		\begin{align}
			(\mathtt{P2.2}):~&\min\limits_{\{\boldsymbol{q}[n]\},\{P_k[n]\geq0\}}~\sum\limits_{n=1}^{\tilde N}\mathbbm{1}\bigg({\tt SNR}(\boldsymbol{q}[n],\{P_k[n]\})\bigg)\nonumber\\
			\mathrm{s.t.}~&\frac{1}{\tilde{N}}\sum\limits_{n=1}^{\tilde{N}} P_k[n]\leq P^{\rm ave}_k, \forall k\in \mathcal{K}\label{08061004}\\
			&\|\boldsymbol{q}[n]-\boldsymbol{q}[n-1]\|^2\leq V^2_{\rm max}{\tilde{\delta}}^2, \forall n\in \tilde{\mathcal N}\label{08061005}\\
			&\boldsymbol{q}[0]=\boldsymbol{q}_{\rm I}, \boldsymbol{q}[\tilde{N}]=\boldsymbol{q}_{\rm F}.\label{080610055}
		\end{align}
	\end{subequations}
	Problem $(\mathtt{P2.2})$ is still non-convex. To tackle this issue, let $l_n(\boldsymbol{q}[n],\{P_k[n]\})={\tt SNR}(\boldsymbol{q}[n],\{P_k[n]\})-\gamma_{\rm min},\forall n\in \mathcal{\tilde{N}}$ and $\boldsymbol{l}(\{\boldsymbol{q}[n],P_k[n]\})=[l_1(\boldsymbol{q}[1],\{P_k[1]\}),\ldots,l_{\tilde{N}}(\boldsymbol{q}[\tilde{N}],\{P_k[\tilde{N}]\})]$. As a result, problem $(\mathtt{P2.2})$ is equivalently expressed as
	\begin{align}
		(\mathtt{P2.3}):~\min\limits_{\{\boldsymbol{q}[n]\},\{P_k[n]\geq0\}}~&\|\boldsymbol l(\{\boldsymbol{q}[n]\},\{P_k[n]\})\|_0\nonumber\\
		\mathrm{s.t.}~&(\ref{08061004}),(\ref{08061005}), \text{and} ~(\ref{080610055}).\nonumber
	\end{align}
	To handle the zero-norm function in problem $\mathtt{(P2.3)}$, we use $\|\boldsymbol l(\{\boldsymbol{q}[n]\},\{P_k[n]\})\|_1$ to approximate $\|\boldsymbol l(\{\boldsymbol{q}[n]\},\{P_k[n]\})\|_0$ \cite{l0}. Note that to reduce the outage probability with minimized energy consumption, the received SNR of each time slot should not be larger than $\gamma_{\min}$. Thus, we have the following constraints:
	${\tt SNR}(\boldsymbol{q}[n],\{P_k[n]\})\leq \gamma_{\rm min},\forall n\in \mathcal{\tilde{N}}$.
	Similar as in problem $\mathtt{(P1.2)}$, we introduce two sets of auxiliary variables $\{a_k[n]\}$ and $\{A_k[n]\}$, $k\in\mathcal K,n\in\mathcal {\tilde{N}}$, and problem $\mathtt{(P2.3)}$ is approximated as
	\begin{subequations}
		\begin{align}
			&(\mathtt{P2.4}):~\max\limits_{\{\boldsymbol{q}[n]\},\{P_k[n]\geq0\},\{A[n]\},\{a_k[n]\}}~\frac{1}{\tilde{N}}\sum\limits_{n=1}^{\tilde{N}} A[n]/\sigma^2\nonumber\\
			&\mathrm{s.t.}~A[n]\le\left(\sum_{k=1}^K a_k[n]\right)^2,\forall n\in\mathcal {\tilde{N}} \label{12171640}\\
			&a_k[n]\le \sqrt{\frac{P_k[n]\beta_0}{(\|\boldsymbol{q}[n]-\boldsymbol{s}_k\|^2+H^2)^{\alpha/2}}},\forall k\in\mathcal K,n\in\mathcal {\tilde{N}} \label{eqn:17b}\\
			&A[n]/\sigma^2\le \gamma_{\min},\forall n\in\mathcal {\tilde{N}}\\
			&(\ref{08061004}),~(\ref{08061005}),~\text{and}~(\ref{080610055}).\nonumber
		\end{align}
	\end{subequations}
	Problem $(\mathtt{P2.4})$ is still non-convex due to non-convex constraints in (\ref{12171640}) and (\ref{eqn:17b}). Specifically, we solve problem $\mathtt{(P2.4)}$ by optimizing the UAV trajectory and GNs' power allocation in an alternating manner via SCA techniques. By applying the similar lower bounds in (\ref{06191622}) and (\ref{06191623}), we can obtain an efficient solution, which is omitted for brevity. Let $\{\boldsymbol{q}^{*}[n]\}$ and $\{P^*_k[n]\}$ denote the obtained trajectory and power allocation, respectively.
	
	{\it Complexity Analysis:} Similar to problem $\mathtt{(P1.3)}$, the complexity of the proposed alternating optimization-based approach for solving problem $\mathtt{(P2.4)}$ includes the complexities of solving the approximated power allocation sub-problem and trajectory design sub-problem in an iterative manner, denoted as sub-problem $\mathtt{(P2.4.1)}$ and sub-problem $\mathtt{(P2.4.2)}$, respectively. Hence, the complexity of problem $\mathtt{(P2.4)}$ can be represented as $\mathbb{O}(L'_{\rm alt}(L'_{\rm SCA1}+L'_{\rm SCA2})K^{3.5}\tilde{N}^{3.5})$, where $L'_{\rm alt}$, $L'_{\rm SCA1}$, and $L'_{\rm SCA2}$ are the iteration numbers of alternating optimization for problem $\mathtt{(P2.4)}$, SCA for sub-problem $\mathtt{(P2.4.1)}$, and SCA for sub-problem $\mathtt{(P2.4.2)}$, respectively.
	
	Finally, we use an additional step to obtain the GNs' power allocation $\{P_k[n]\}$ for problem $\mathtt{(P2.2)}$ under the obtained UAV trajectory $\{\boldsymbol{q}^{*}[n]\}$, for which the problem is given as
	\begin{align}
		(\mathtt{P2.5}):~\min\limits_{\{P_k[n]\geq0\}}~&\sum\limits_{n=1}^{\tilde{N}}\mathbbm{1}({\tt SNR}(\boldsymbol{q}^{*}[n],\{P_k[n]\}))\nonumber\\
		\mathrm{s.t.}~&\frac{1}{\tilde{N}}\sum\limits_{n=1}^{\tilde{N}}P_k[n]\leq P^{\rm ave}_k, \forall k\in \mathcal{K}.\nonumber
	\end{align}
	To solve problem $\mathtt{(P2.5)}$, we sort the time slots based on the SNR $\{{\tt SNR}(\boldsymbol{q}^{*}[n],\{P^{*}_k[n]\})\}$, i.e., ${\tt SNR}(\boldsymbol{q}^{*}[\pi(1)],\{P^{*}_k[\pi(1)]\}) \ge \cdots\ge {\tt SNR}(\boldsymbol{q}^{*}[\pi(\tilde{N})],\{P^{*}_k[\pi(\tilde{N})]\})$, in which $\pi(\cdot)$ denotes the permutation over $\mathcal {\tilde{N}}$. Then, we allocate the GNs' transmit power over a subset $\mathcal{N'}$ of time slots with the highest SNR values, i.e., $\mathcal N' = \{\pi(1),\ldots, \pi(N')\}$, where $N'$ is a variable to be determined. To find $N'$ and the corresponding power allocation, we solve the following feasibility problem.
	\begin{subequations}\label{1112}
		\begin{align}
			&(\mathtt{P2.6}):{\rm find}~~{\{P_k[n]\geq0\}}, \forall n\in \mathcal{N'}, k\in \mathcal{K}\nonumber\\
			&\mathrm{s.t.}~{\tt SNR}(\boldsymbol{q}^{*}[\pi(n)],\{P_k[\pi(n)]\})\geq \gamma_{\rm min}, \forall {\pi(n)}\in \mathcal{N'}\label{1110}\\
			&~~~~~\frac{1}{N'}\sum\limits_{n=1}^{N'}P_k[{\pi(n)}]\leq P^{\rm ave}_k,  \forall k\in \mathcal{K}.
		\end{align}
	\end{subequations}
By letting $\rho'_k[n]=\sqrt{P_k}[n]$, problem $(\mathtt{P2.6})$ can be transformed into a convex form and thus be solved optimally via CVX \cite{2}. By solving problem $(\mathtt{P2.6})$ under given $N'$ together with a bisection search over $\tilde{\mathcal{N}}$, we can find a high-quality solution to problem $(\mathtt{P2.5})$. By combining this together with $\{\boldsymbol{q}^*[n]\}$, an efficient solution of $N'$ and the corresponding power allocation at GNs to problem $(\mathtt{P2})$ is finally obtained.
	
	In addition, in order to guarantee the performance of the obtained solution to problem $\mathtt{(P2)}$, we adopt similar trajectory designs presented in {\it Remark \ref{initial_trajectory}}, and choose the one with the best performance as the initial point.
	
	Since the problem formulations and solutions of the two original problems are very similar, we summarize the similarities and differences in Table \ref{diff} as the following for the sake of clarity.
	\begin{table*}[!t]
		\centering
			\caption{Similarities and Differences Between the Solutions}\label{diff}
			\begin{tabular}{|m{40pt}|m{450pt}|}
				\hline
				Similarities&1)	We first obtain the optimal solutions to the relaxed problems of $(\mathtt{P1})$ and $(\mathtt{P2})$ in the special case with infinite $T$ to obtain key engineering insights by solving the dual problems and time-sharing problems.

				2) We propose an alternating-optimization-based algorithm to obtain efficient solutions to the original problems $(\mathtt{P1})$ and $(\mathtt{P2})$ under any finite $T$, whose initial points are based on the optimal solutions under the special case.

				3) At each iteration of the alternating optimization, we use the SCA approach to approximate the non-convex problems as convex optimization problems.\\
				\hline
				Differences&1) As for problem $(\mathtt{P1})$, the objective function of the alternating-optimization-based algorithm is the same as the original problem $(\mathtt{P1})$. Therefore, after the alternation, the efficient solution to problem $(\mathtt{P1})$ is directly obtained.
				
				2) As for problem $(\mathtt{P2})$, before performing the alternating-optimization-based algorithm, we approximate the outage probability minimization problem $(\mathtt{P2})$ to the average SNR maximization problem $(\mathtt{P2.4})$. After the alternation, we obtain an efficient trajectory and then use a bisection method over the time slots according to the SNR order to find the power allocation for the GNs. Therefore, the efficient solution to problem $(\mathtt{P2})$ is finally obtained.

				3) As for problem (P1), the GNs transmit their messages based on the water-filling-like power allocation over time. While for problem (P2), the GNs adopt an on-off power allocation over time, where the GNs may remain silent in the outage status when the wireless channels become bad, such that the transmit power can be reserved for non-outage transmission at other time instants.\\
				\hline
		\end{tabular}
	\end{table*}

	\section{Numerical Results}\label{s7}	
	In the simulation, we consider the scenario with $10$ GNs, which are located at $(20,10)$~m, $(30,28)$~m, $(46,0)$~m, $(56,24)$~m, $(94,168)$~m, $(100,200)$~m, $(112,176)$~m, $(162,0)$~m, $(178,40)$~m, and $(200,6)$~m. We set $\beta_0=-30$ dB, $\sigma^2 = -60$ dBm, $K =10$, $\alpha = 2.8$, $V_{\rm max}=40$ m/s, $N = \tilde{N} =128$, $H = 50$ m, $\boldsymbol{q}_{\rm I}=$ ($0,0$) m, $\boldsymbol{q}_{\rm F}= $ ($200,200$)~m, and $\gamma_{\rm min} = 27.4$ dB, unless otherwise stated.
	
	For each scenario, we first show the system setup and the obtained trajectories under given $T$.
	Next, we compare the performance of our proposed design versus the following scheme together with the three designs presented in {\it Remark \ref{initial_trajectory}}.
	\begin{itemize}
		\item \textbf{Trajectory design only}: In this scheme, the GNs use the uniform power allocation and accordingly the UAV's trajectories are obtained by solving the trajectory optimization problems in $(\mathtt{P1.3})$ and $(\mathtt{P2.4})$, respectively.
	\end{itemize}

	\subsection{Rate Maximization in Delay-Tolerant Scenario}
	\begin{figure}[!h]
		\centering
    \includegraphics[width=8cm]{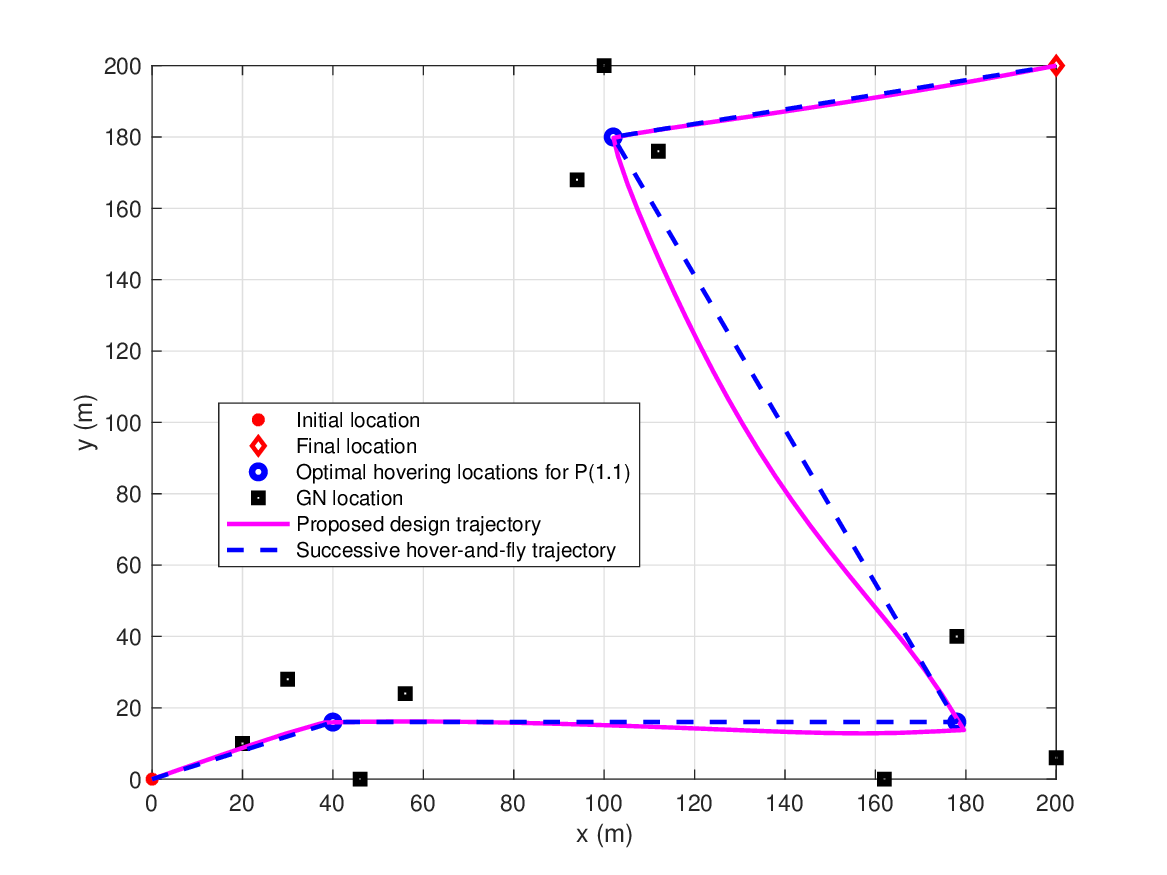}
		\caption{System setup and the obtained trajectories with $T=20$~s in the delay-tolerant scenario.}\label{fig:trajectory_tolerant}
	\end{figure}
	Fig. \ref{fig:trajectory_tolerant} shows the system setup and the obtained trajectories for problem $(\mathtt{P1})$ with $T=20$~s.
	It is observed that there are ${V}=3$ optimal hovering locations for problem $\mathtt{(P1.1)}$.	
	
	\begin{figure}[!h]
		\centering
    \includegraphics[width=8cm,height=6.5cm]{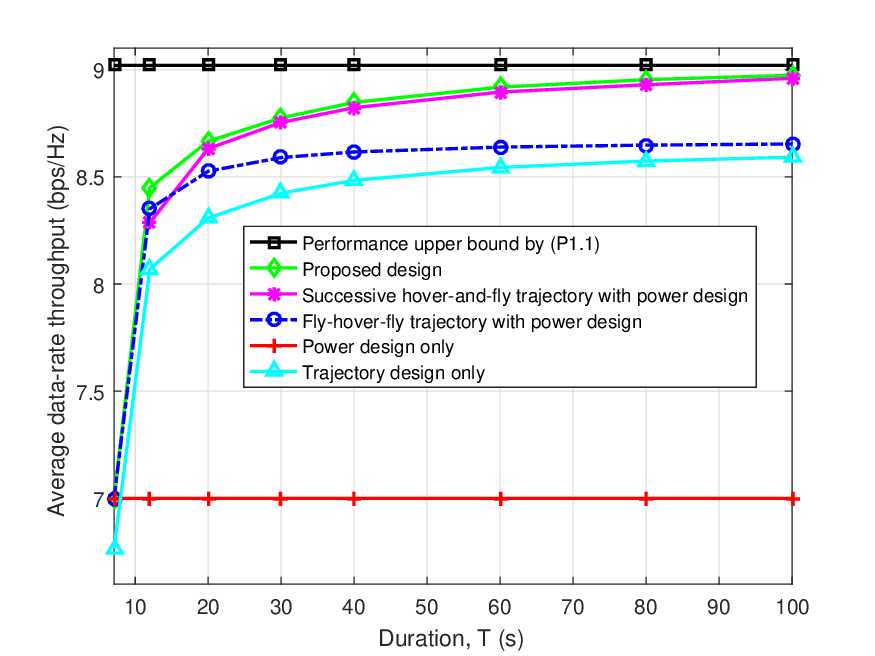}
			\caption{Average data-rate throughput versus the flight duration $T$ in the delay-tolerant scenario.}\label{fig:rate_tolerant}
	\end{figure}	
	Fig. \ref{fig:rate_tolerant} shows the average data-rate throughput of the system versus the flight duration $T$, where $P^{\rm ave}_k=30$ dBm, $\forall k\in \mathcal K$.
	It is observed that the proposed design achieves higher average data-rate throughput than the other benchmark schemes. Furthermore, with sufficiently large $T$, the proposed design is observed to approach the performance upper bound achieved by problem $\mathtt{(P1.1)}$ with the UAV's flight speed constraints ignored. The successive hover-and-fly trajectory with power design is observed to perform close to the proposed design, which shows the significance of the optimized hovering locations.
	
	\begin{figure}[!h]
		\centering
    \includegraphics[width=8cm,height=6.5cm]{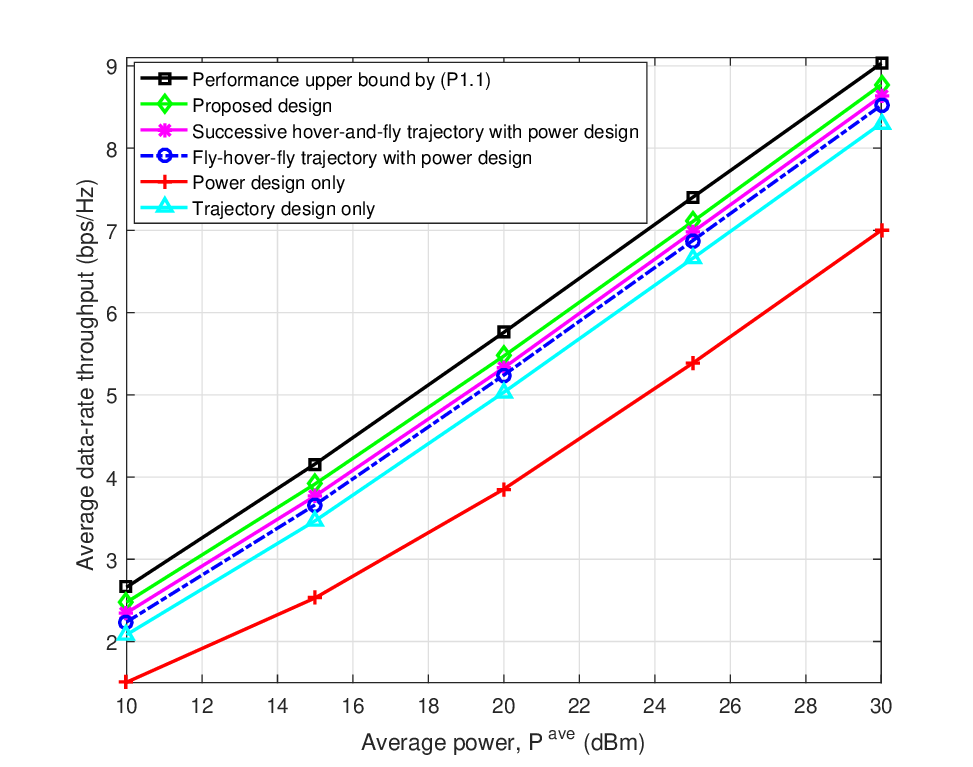}
			\caption{Average data-rate throughput versus the GN's maximum average power constraint $P^{\rm ave}$ in the delay-tolerant scenario.}\label{fig:power_tolerant}
	\end{figure}	
	Fig. \ref{fig:power_tolerant} shows the average data-rate throughput of the system versus the GN's maximum average power $P^{\rm ave}_k=P^{\rm ave}$, $\forall k\in \mathcal K$, where $T = 20$ s.
	It is observed that as $P^{\rm ave}$ increases, the average data-rate throughputs of all the methods increase. Similar observations are made as in Fig. \ref{fig:rate_tolerant}.
	
	\subsection{Outage Probability Minimization in Delay-Sensitive Scenario}	
	\begin{figure}[!h]
		\centering
    \includegraphics[width=8cm]{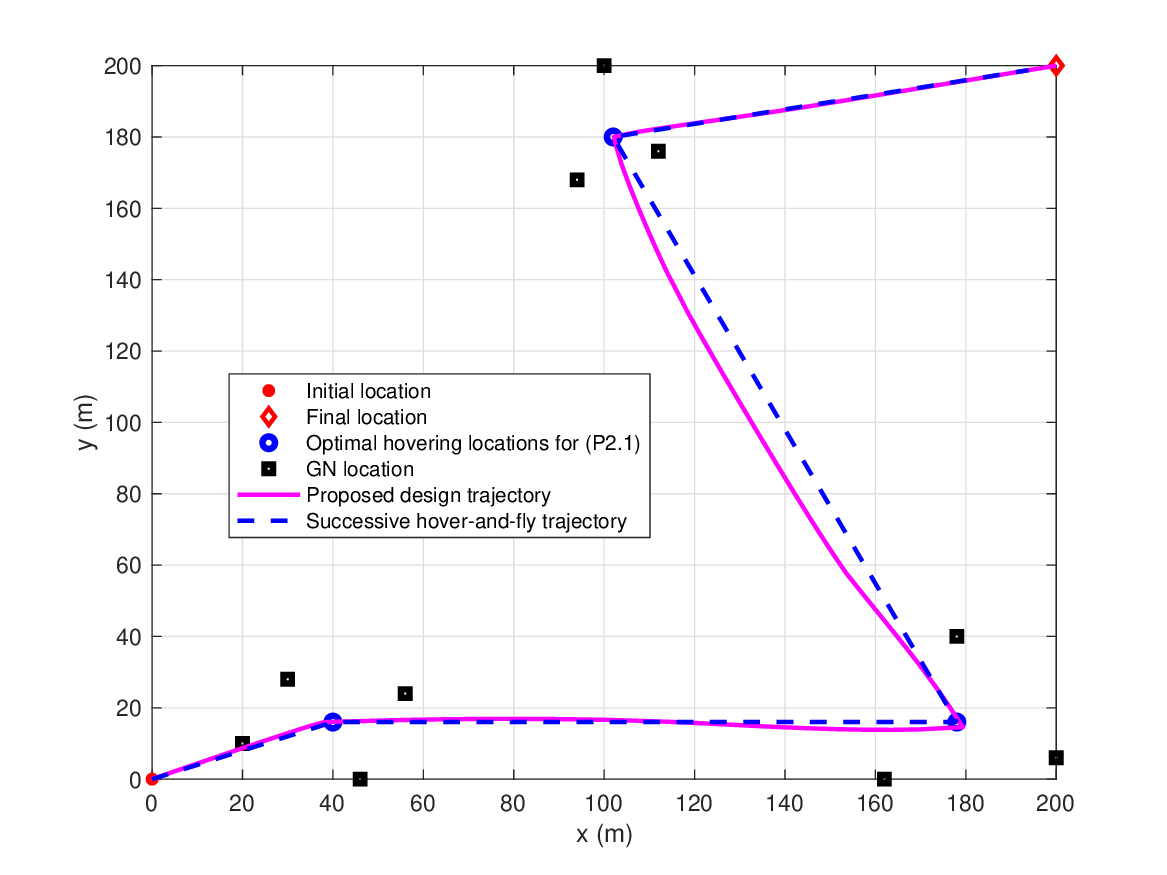}
		\caption{System setup and the obtained trajectories with $T=20$~s in the delay-sensitive scenario.}\label{fig:trajectory_sensitive}
	\end{figure}	
	Fig. \ref{fig:trajectory_sensitive} shows the system setup and the obtained trajectories with $T=20$~s for problem $(\mathtt{P2})$.
	The optimized hovering locations are observed to be the same as those in Fig. \ref{fig:trajectory_tolerant} under this setup, which is consistent with our analysis in Section \ref{s5}. The obtained trajectories are observed to be similar to those in Fig. \ref{fig:trajectory_tolerant}, with only a slight difference. However, the strategies of the UAV's time schedules and GNs' power allocations are much different for the two problems. For example, at the proposed design trajectories, according to the simulation results, the UAV hovers over the first optimal hovering location for $4.38$ s in Fig. \ref{fig:trajectory_tolerant}, while for $4.69$ s in Fig. \ref{fig:trajectory_sensitive}. Moreover, Table \ref{t3} shows the GNs' transmit powers in dBm at the first optimal hovering location of the two problems, which indicate the different power allocation strategies.
		
		\begin{table*}[!t]
			\centering
				\caption{GNs' Transmit Power at the First Optimal Hovering Location ({\upshape d{B}m})}
				\begin{tabular}{|c|c|c|c|c|c|c|c|c|c|c|}
					\hline
					~&GN 1&GN 2&GN 3&GN 4&GN 5&GN 6&GN 7&GN 8&GN 9&GN 10\\
					\hline
					Fig. \ref{fig:trajectory_tolerant}&$34.0$&$33.9$&$33.7$&$33.2$&$19.0$&$17.7$&$17.5$&$22.8$&$21.0$&$21.3$\\
					\hline
					Fig. \ref{fig:trajectory_sensitive}&$33.5
					$&$33.5$&$33.0$&$32.5$&$19.4$&$18.3$&$18.0$&$21.1$&$19.8$&$19.9$\\
					\hline
			\end{tabular}\label{t3}
		\end{table*}

	\begin{figure}[!h]
		\centering
    \includegraphics[width=8cm,height=6.5cm]{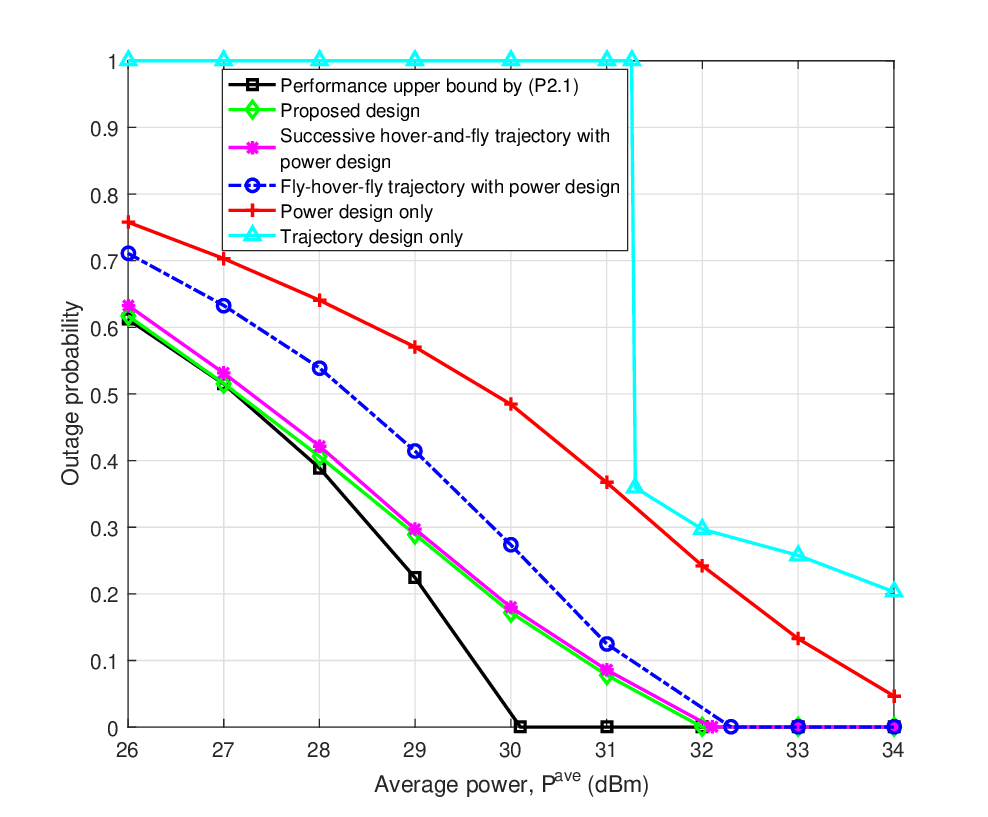}
			\caption{Outage probability versus the GN's maximum average power constraint $P^{\rm ave}$ in the delay-sensitive scenario.}\label{fig:power_sensitive}
	\end{figure}	
	Fig. \ref{fig:power_sensitive} shows the outage probability of the system versus the GN's maximum average power $P^{\rm ave}_k=P^{\rm ave}$, $\forall k\in \mathcal K$, where $T=20$~s. It is observed that when $P^{\rm ave}$ is less than $31.3$ dBm, the outage probability achieved by the trajectory design only scheme is $1$; while that achieved by other schemes is less than $1$.
It is because that with the constant uniform power allocation for the trajectory design only scheme, the received SNR at the UAV only depends on the channel condition, i.e., the UAV trajectory.
Hence, it is always outage until the average power is up to 31.3 dBm, which meets the requirement of the location with the best channel condition. Consequently, there is a sharp decrease for the trajectory design only scheme. This shows that power optimization is quite significant in this case.
It is also observed that the performance gap between the proposed method and the upper bound increases as the average power increases.
This is intuitive, as the upper bound, the UAV always hovers at the optimal locations to enjoy the strong channel condition and achieve the best communication performance, while for the proposed method, the power is prior to allocating to the time duration at the optimal locations and then for the communication in the path connecting the optimal locations.
Furthermore, it is observed that our proposed design considerably outperforms other benchmark schemes in all regions of transmit power, by jointly designing the UAV's trajectory and the GNs' power allocation.

	\begin{figure}[!h]
		\centering
    \includegraphics[width=8cm,height=6.5cm]{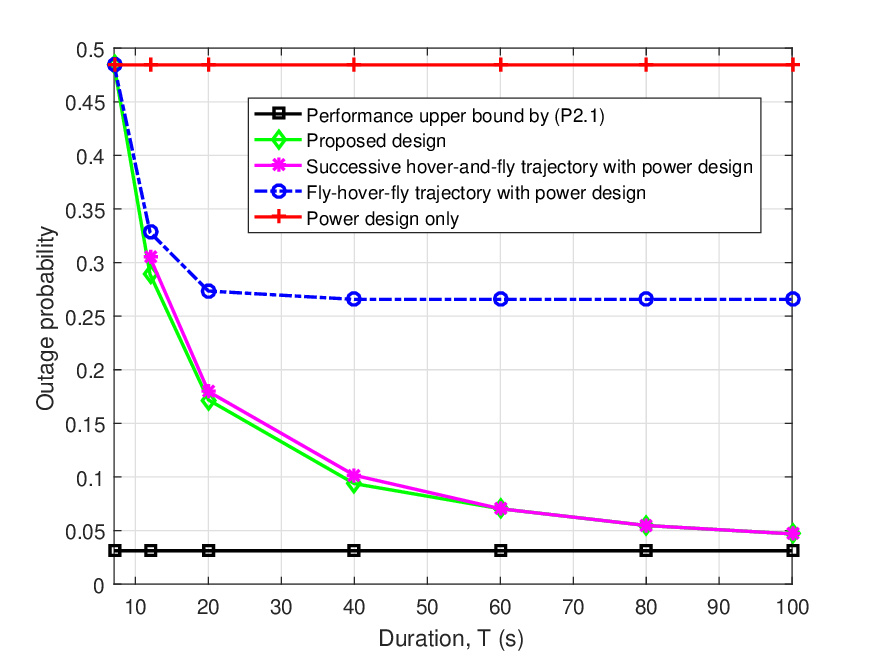}
			\caption{Outage probability versus the flight duration $T$ in the delay-sensitive scenario.}\label{fig:outage_sensitive}
	\end{figure}	
	Fig. \ref{fig:outage_sensitive} shows the outage probability versus the flight duration $T$, where $P^{\rm ave}_k=30$ dBm, $\forall k\in \mathcal K$. Notice that the trajectory design only scheme always leads to the outage probability of one, and therefore, this scheme is not shown in this figure. It is observed that the proposed design achieves a lower outage probability than other benchmark schemes. Furthermore, with sufficiently large $T$, the proposed design leads to similar performance as the performance upper bound achieved by problem $\mathtt{(P2.1)}$ with the UAV's flight speed constraints ignored.
	
	\section{Conclusion}\label{s11}
	
	In this paper, we considered the UAV-enabled data collection from multiple GNs with distributed beamforming. We maximized the average data-rate throughput and minimized the transmission outage probability, by jointly optimizing the UAV's trajectory and the GNs' power allocation over time. To deal with these challenging problems, we first optimally solved the relaxed problem without considering the UAV's flight speed constraints. The optimal solutions indicated that the UAV should successively hover over the same location set for both problems, but with different power allocation strategies. Next, we used the techniques from convex optimization and approximation to find the sub-optimal solutions to the general problems. Finally, we conducted simulations to show the effectiveness of our proposed design. As the space is limited, there are still several important issues unaddressed in this paper. These issues will be briefly discussed in the following to motivate future work.

	\begin{itemize}
		\item When an imperfect location model is considered, our proposed schemes are extendable by e.g., considering a bounded location error model with a given estimation error, which is generally handled by the robust optimization techniques \cite{ZhongSecure2019}. However, it is non-trivial to solve the resultant robust optimization problem with the estimation errors taken into account. Thus, we would like to leave this interesting topic for our future work.
		
		\item Under mixed LoS and non-LoS cases (e.g., probabilistic LoS models \cite{3GPP}), 3D trajectory design is beneficial since there is a tradeoff between the link distance and LoS probability. In particular, the higher altitude means the longer link distance, but also leads to higher LoS probability. Hence, it generally exists an optimal UAV altitude. How to optimize the problems in 3D under other channel models (e.g., probabilistic model) is another interesting topic that requires further investigation.
		
		\item According to the estimation schemes in \cite{ZengOptimized2015}, the GNs may only know imperfect channel state information (CSI) due to the errors of channel estimation and channel quantization. As a result, properly designing the CSI acquisition under our setup to balance the acquisition overhead and the CSI accuracy is a significant work. Furthermore, how to optimize the distributed beamforming under such imperfect CSI (e.g., via robust optimization techniques \cite{VorobyovRobust2003}) is another issue that requires further investigation.
		
		\item In order to perform the distributed beamforming, we need to synchronize the data in advance. However, due to the random errors such as quantization noise and wireless channel distortion, the data synchronization scheme is a non-trivial task and beyond the scope of the paper, thus deserving a dedicated study. Hence, we would like to leave this interesting and important topic as our future work.
		
		\item Finally, multiple UAVs and the received beamforming at the multi-antenna UAVs, which are beneficial for reliable and fast data collection, can be taken into consideration. Nevertheless, there have been various practical issues to be addressed. How to extend our proposed results in these scenarios is a challenge future direction.
		
	\end{itemize}


\begin{thebibliography}{1}
		\bibliographystyle{IEEEbib}
		
		\bibitem{FengOutage2020}
		T. Feng, L. Xie, J. Yao, and J. Xu, ``Outage probability minimization for UAV-enabled
		data collection with distributed beamforming," in {\it Proc. IEEE Int. Conf. Commun. Workshops (ICC Workshops)}, Virtual Conference, Jun. 2020, pp. 1-6.				
		
		\bibitem{MozaffariBeyond2019}
		M. Mozaffari, A. T. Z. Kasgari, W. Saad, M. Bennis, and M. Debbah,
		``Beyond 5G with UAVs: Foundations of a 3D wireless cellular network,"
		{\it IEEE Trans. Wireless Commun.}, vol. 18, no. 1, pp. 357-372, Jan. 2019.
		
		\bibitem{LiPlacement2018}
		P. Li and J. Xu, ``Placement optimization for UAV-enabled wireless networks with multi-hop backhauls," {\it J. Commun. Inf. Netw.}, vol. 3, no. 4, pp. 64-73, Dec. 2018.
		
		\bibitem{Alzenad3-D2017}
		M. Alzenad, A. El-Keyi, F. Lagum, and H. Yanikomeroglu, ``3-D placement of an unmanned aerial vehicle base station (UAV-BS) for energy-efficient maximal coverage," {\it IEEE Wireless Commun. Lett.}, vol. 6, no. 4, pp. 434-437, Aug. 2017.
		
		\bibitem{ZhongSecure2019}
		C. Zhong, J. Yao, and J. Xu, ``Secure UAV communication with cooperative jamming and trajectory control," {\it IEEE Commun. Lett.}, vol. 23, no. 2, pp. 286-289, Feb. 2019.
		
		\bibitem{YaoSecrecy2019}
		J. Yao and J. Xu, ``Secrecy transmission in large-scale UAV-enabled wireless networks," {\it IEEE Trans. Commun}, vol. 67, no. 11, pp. 7656-7671, Nov. 2019.


        \bibitem{BanagarPerformance2020}
			M. Banagar and H. S. Dhillon, ``Performance characterization of canonical mobility models in drone cellular networks," {\it IEEE Trans. Wireless Commun.}, vol. 19, no. 7, pp. 4994-5009, Jul. 2020.

            \bibitem{SharmaRandom2019}
			P. K. Sharma and D. I. Kim, ``Random 3D mobile UAV networks: Mobility modeling and coverage probability," {\it IEEE Trans. Wireless Commun.}, vol. 18, no. 5, pp. 2527-2538, May 2019.

            \bibitem{AmerMobility2020}
			R. Amer, W. Saad, and N. Marchetti, ``Mobility in the sky: Performance and mobility analysis for cellular-connected UAVs," {\it IEEE Trans. Commun.}, vol. 68, no. 5, pp. 3229-3246, May 2020.

		
		\bibitem{ZhangTrajectory2019}
		S. Zhang and R. Zhang, ``Trajectory optimization for cellular-connected UAV under outage duration constraint," {\it J. Commun. Inf. Netw.}, vol. 4, no. 4, pp. 55-71, Dec. 2019.
		
		\bibitem{XieThroughput2019}
		L. Xie, J. Xu, and R. Zhang, ``Throughput maximization for UAV-enabled wireless powered communication networks," {\it IEEE Internet Things J.}, vol. 6, no. 2, pp. 1690-1703, Apr. 2019.
		
		\bibitem{XuUAV2018}
		J. Xu, Y. Zeng, and R. Zhang, ``UAV-enabled wireless power transfer: Trajectory
		design and energy optimization," {\it IEEE Trans. Wireless Commun.}, vol. 17, no. 8, pp. 5092-5106, Aug. 2018.
		
		\bibitem{XieCommon2020}
		L. Xie, J. Xu, and Y. Zeng, ``Common throughput maximization for UAV-enabled interference channel with wireless powered communications," {\it IEEE Trans. Commun.}, vol. 68, no. 5, pp. 3197-3212, May 2020.
		
		\bibitem{ChenEfficient2020}
		J. Chen and D. Gesbert, ``Efficient local map search algorithms for the placement of flying relays," {\it IEEE Trans. Wireless Commun.}, vol. 19, no. 2, pp. 1305-1319, Feb. 2020.
		
		\bibitem{ChenMultiple2018}
		Y. Chen, N. Zhao, Z. Ding, and M. Alouini, ``Multiple UAVs as relays: Multi-hop single link versus multiple dual-hop links," {\it IEEE Trans. Wireless Commun.}, vol. 17, no. 9, pp. 6348-6359, Sep. 2018.		
		
		\bibitem{ZhouMobile2020}
		F. Zhou, R. Q. Hu, Z. Li, and Y. Wang, ``Mobile edge computing in unmanned aerial vehicle networks," {\it IEEE Wireless Commun.}, vol. 27, no. 1, pp. 140-146, Feb. 2020.
		
		\bibitem{HuJoint2019}
		Q. Hu, Y. Cai, G. Yu, Z. Qin, M. Zhao, and G. Y. Li, ``Joint offloading and trajectory design for UAV-enabled mobile edge computing systems," {\it IEEE Internet Things J.}, vol. 6, no. 2, pp. 1879-1892, Apr. 2019.
		
		\bibitem{LiEnergy2019}
		K. Li, R. C. Voicu, S. S. Kanhere, W. Ni, and E. Tovar, ``Energy efficient legitimate wireless surveillance of UAV communications," {\it IEEE Trans. Veh. Technol.}, vol. 68, no. 3, pp. 2283-2293, Mar. 2019.
		
		\bibitem{lei}
		M. Li, Y. Liu, and L. Chen, ``Nonthreshold-based event detection for 3D environment monitoring in sensor networks," {\it IEEE Trans. Knowl. Data Eng.}, vol. 20, no. 12, pp. 1699-1711, Dec. 2008.
		
\bibitem{Jawad}
H. Jawad, R. Nordin, S. Gharghan, A. Jawad, and M. Ismail, ``Energy-efficient wireless sensor networks for precision agriculture: A review," {\it Sensors}, vol. 17, no. 8, pp. 1781, Feb. 2017.

\bibitem{Nellore}
K. Nellore and G. Hancke, ``A survey on urban traffic management system using wireless sensor networks," {\it Sensors}, vol. 16, no. 2, pp. 157, Jan. 2016.
		
		
			\bibitem{AlmasoudEnergy2018}
			A. M. Almasoud, M. Y. Selim, A. Alqasir, T. Shabnam, A. Masadeh, and A. E. Kamal, ``Energy efficient data forwarding in disconnected networks using cooperative UAVs," in {\it Proc. IEEE GLOBECOM}, Abu Dhabi, United Arab Emirates, Dec. 2018, pp. 1-6.
				
		\bibitem{GongFlight2018} J. Gong, T. Chang, C. Shen, and X. Chen, ``Flight time minimization
		of UAV for data collection over wireless sensor networks," {\it IEEE J. Sel.
			Areas Commun.}, vol. 36, no. 9, pp. 1942-1954, Sep. 2018.
		
		\bibitem{LiJoint2019}
		J. Li, H. Zhao, H. Wang, F. Gu, J. Wei, H. Yin, and B. Ren, ``Joint
		optimization on trajectory, altitude, velocity and link scheduling for
		minimum mission time in UAV-aided data collection," {\it IEEE Internet
			Things J.}, vol. 7, no. 2, pp. 1464-1475, Feb. 2020.
		
		\bibitem{WangEnergy2019} Z. Wang, R. Liu, Q. Liu, J. S. Thompson, and M. Kadoch, ``Energy
		efficient data collection and device positioning in UAV-assisted IoT,"
		{\it IEEE Internet Things J.}, vol. 7, no. 2, pp. 1122-1139, Feb. 2020.
		
		\bibitem{ZhanEnergy2018} C. Zhan, Y. Zeng, and R. Zhang, ``Energy-efficient data collection in
		UAV enabled wireless sensor network," {\it IEEE Wireless Commun. Lett.},
		vol. 7, no. 3, pp. 328-331, Jun. 2018.
		
		\bibitem{ZhangMulti2020}
		J. Zhang, Y. Zeng, and R. Zhang, ``Multi-antenna UAV data harvesting: Joint trajectory and communication optimization," {\it J. Commun. Inf. Netw.}, vol. 5, no. 1, pp. 86-99, Mar. 2020.
		
		\bibitem{You3D2019} C. You and R. Zhang, ``3D trajectory optimization in Rician fading for
		UAV-enabled data harvesting," {\it IEEE Trans. Wireless Commun.}, vol. 18,
		no. 6, pp. 3192-3207, Jun. 2019.
		
		\bibitem{Li2020} P. Li and J. Xu, ``Fundamental rate limits of UAV-enabled multiple
		access channel with trajectory optimization," {\it IEEE Trans. Wireless Commun.}, vol. 19, no. 1, pp. 458-474, Jan. 2020.
		
		\bibitem{1}
		R. Mudumbai, D. R. Brown, U. Madhow, and H. V. Poor, ``Distributed transmit beamforming: Challenges and recent progress," {\it IEEE Commun. Mag.}, vol. 47, no. 2, pp. 102-110, Feb. 2009.
		
		\bibitem{Z}
		J. Xu, Z. Zhong, and B. Ai, ``Wireless powered sensor networks: Collaborative energy beamforming considering sensing and circuit power consumption,"
		{\it IEEE Wireless Commun. Lett.}, vol. 5, no. 4, pp. 344-347, Aug. 2016.
		
		\bibitem{Haro}
		B. B. Haro, S. Zazo, and D. P. Palomar, ``Energy efficient collaborative beamforming in wireless sensor networks," {\it IEEE Trans. Signal Process.}, vol. 62, no. 2, pp. 496-510, Jan. 2014.

		\bibitem{Barriac}
		R. Mudumbai, G. Barriac, and U. Madhow, ``On the feasibility of distributed beamforming in wireless networks," {\it IEEE Trans. Wireless Commun.}, vol. 6, no. 5, pp. 1754-1763, May 2007.
		
		\bibitem{m2}
		D. R. Brown III and H. V. Poor, ``Time-slotted round-trip
		carrier synchronization for distributed beamforming," {\it IEEE Trans. Signal Proc.}, vol. 56, no. 11, pp. 5630-5643, Nov. 2008.
		
		
			\bibitem{ShangUnmanned2019}
			B. Shang, L. Liu, J. Ma, and P. Fan, ``Unmanned aerial vehicle meets vehicle-to-everything in secure communications," {\it IEEE Commun. Mag.}, vol. 57, no. 10, pp. 98-103, Oct. 2019.
			
			\bibitem{LinThe2018}
			X. Lin, V. Yajnanarayana, S. D. Muruganathan, S. Gao, H. Asplund,
			H. Maattanen, M. Bergstrom, S. Euler, and Y. E. Wang, ``The sky is
			not the limit: LTE for unmanned aerial vehicles," {\it IEEE Commun. Mag.},
			vol. 56, no. 4, pp. 204-210, Apr. 2018.
			
			\bibitem{3GPP}
			3GPP TR 36.777, ``Study on enhanced LTE support for aerial vehicles,"
			Dec. 2017.
		
		
			\bibitem{DabiriAnalytical2020}
			M. T. Dabiri, H. Safi, S. Parsaeefard, and W. Saad, ``Analytical channel models for millimeter wave UAV networks under hovering fluctuations," {\it IEEE Trans. Wireless Commun.}, vol. 19, no. 4, pp. 2868-2883, Apr. 2020.
			
			
			\bibitem{BanagarImpact}
			M. Banagar, H. S. Dhillon, and A. F. Molisch, ``Impact of UAV jittering on the air-to-ground wireless channel," [Online]. Available: \url{https://arxiv.org/abs/2004.02771}
		
		
		
			\bibitem{ZengOptimized2015}
			Y. Zeng and R. Zhang, ``Optimized training design for wireless energy transfer," {\it IEEE Trans. Commun.}, vol. 63, no. 2, pp. 536-550, Feb. 2015.
				
		\bibitem{yu}
		W. Yu and R. Lui, ``Dual methods for nonconvex spectrum
		optimization of multicarrier systems," {\it IEEE Trans. Commun.}, vol. 54, no. 7, pp. 1310-1322, Jun. 2006.
		
		\bibitem{2}
		S. Boyd and L. Vandenberghe, {\it Convex Optimization}, Cambidge Univ.
		Press, 2004.
		
		\bibitem{ell}
		S. Boyd, ``Convex optimization II," Stanford University. [Online]. Available: \url{http://www.stanford.edu/class/ee364b/lectures}
		
			\bibitem{RanOn2002}
			R. Ran, ``On the complexity of matrix product", {\it SIAM J. Comput.}, vol 32, no. 5, pp. 144-151, Apr. 2002.
			
			\bibitem{WuJoint2018}
			Q. Wu, Y. Zeng, and R. Zhang, ``Joint trajectory and communication
			design for multi-UAV enabled wireless networks," {\it IEEE Trans. Wireless Commun.}, vol. 17, no. 3, pp. 2109-2121, Mar. 2018.
			
			
			\bibitem{ValiulahiMulti2020}
			I. Valiulahi and C. Masouros, ``Multi-UAV deployment for throughput maximization in the presence of co-channel interference,''  {\it IEEE Internet Things J.}, vol. 8, no. 5, pp. 3605-3618, Mar. 2021.
		
		
		\bibitem{l0}
		J. A. Tropp, ``Algorithms for simultaneous sparse approximation. Part II: Convex relaxation," {\it Signal Process.}, vol. 86, no. 3, pp. 589-602, Mar. 2006.
		\bibliography{myreference}
		
		\bibitem{VorobyovRobust2003}
		S. A. Vorobyov, A. B. Gershman, and Z. Luo, ``Robust adaptive beamforming using worst-case performance optimization: A solution to the signal mismatch problem," {\it IEEE Trans. Signal Process.}, vol. 51, no. 2, pp. 313-324, Feb. 2003.
	
	\end{thebibliography}
\end{document}